\documentclass{aa}
\usepackage{rotating}
\usepackage{graphicx}
\usepackage{txfonts}
\begin{document}
\title{A photometric and spectroscopic investigation of star formation  in the very young open cluster NGC\,6383\thanks{Based on observations collected at the European Southern Observatory (ESO, La Silla, Chile).}\fnmsep\thanks{Table 2 is only available in electronic form at the CDS via anonymous ftp to cdsarc.u-strasbg.fr (130.79.128.5) or via http://cdsweb.u-strasbg.fr/cgi-bin/gcat?J/A+A/}}
\date{Received date / Accepted date}
\author{G.\ Rauw\inst{1}\fnmsep\thanks{Research Associate FRS-FNRS (Belgium)}, J.\ Manfroid\inst{1}\fnmsep\thanks{Research Director FRS-FNRS (Belgium)}, \and M.\ De Becker\inst{1}\fnmsep\thanks{Postdoctoral Researcher FRS-FNRS (Belgium)}}
\offprints{G.\ Rauw}
\mail{rauw@astro.ulg.ac.be}
\institute{GAPHE, Institut d'Astrophysique et de G\'eophysique, Universit\'e de Li\`ege, All\'ee du 6 Ao\^ut 17, B\^at B5c, 4000 Li\`ege, Belgium}
\authorrunning{G.\ Rauw et al.\ }
\titlerunning{The very young open cluster NGC\,6383}

\abstract{The very young open cluster NGC\,6383 centered on the O-star binary HD\,159176 is an interesting place for studying the impact of early-type stars with strong radiation fields and powerful winds on the formation processes of low-mass stars.}{To investigate this process, it is necessary to determine the characteristics (age, presence, or absence of circumstellar material) of the population of low-mass pre-main-sequence (PMS) stars in the cluster.}{We obtained deep $U\,B\,V\,(R\,I)_c\,H\alpha$ photometric data of the entire cluster as well as medium-resolution optical spectroscopy of a subsample of X-ray selected objects.}{Our spectroscopic data reveal only very weak H$\alpha$ emission lines in a few X-ray selected PMS candidates. We photometrically identify a number of H$\alpha$ emission candidates but their cluster membership is uncertain. We find that the fainter objects in the field of view have a wide range of extinction (up to $A_V = 20$), one X-ray selected OB star having $A_V \simeq 8$.}{Our investigation uncovers a population of PMS stars in NGC\,6383 that are probably coeval with HD\,159176. In addition, we detect a population of reddened objects that are probably located at different depths within the natal molecular cloud of the cluster. Finally, we identify a rather complex spatial distribution of H$\alpha$ emitters, which is probably indicative of a severe contamination by foreground and background stars.} 
\maketitle
\section{Introduction \label{intro}}
Substantial efforts have been dedicated to the investigation of the low-mass star formation process in very young open clusters harbouring a population of early-type stars. Many of these studies were inspired by X-ray observations of these open clusters (with either {\it Chandra} or {\it XMM-Newton}) that found a wealth of moderately bright X-ray sources associated with low-mass pre-main-sequence (PMS) stars (e.g., Prisinzano et al.\ \cite{Prisinzano}, Sana et al.\ \cite{Sana}). A strong and often variable X-ray emission is frequently observed for PMS stars in the classical and the weak-line T\,Tauri (hereafter cTTs and wTTs respectively) stage. However, enhanced X-ray emission is only one characteristic of PMS stars; in addition to a Li overabundance, position in a colour-magnitude diagram, photometric variability, H$\alpha$ emission, and near-infrared excesses that can help us identify the population of PMS stars in a given cluster. Unfortunately, all these criteria are biased to some extent: X-ray and H$\alpha$ emission are strongly variable and could be weaker over some part of the PMS lifetime or over some stellar mass ranges; the Li lines can only be measured for rather bright objects observable using medium resolution spectroscopy; photometric surveys are heavily affected by foreground or background objects that are unrelated to the cluster, and IR-excesses are only detected for a subset of the PMS objects. Therefore, it is important to consider as many criteria as possible when identifying the stars in the PMS stage.

In this paper, we consider the case of the very young cluster NGC\,6383. The cluster distance was consistently determined to be $1.3 \pm 0.1$\,kpc, while the reddening towards the brighter cluster members corresponds to $E(B - V) = 0.32 \pm 0.02$ (for a review of the cluster properties, see Rauw \& DeBecker \cite{handbook} and references therein). NGC\,6383 is centered on the massive O7((f))\,V + O7((f))\,V binary HD\,159176 (Linder et al.\ \cite{Linder}) and an {\it XMM-Newton} observation detected 76 X-ray sources (Rauw et al.\ \cite{GR03}) in addition to HD\,159176 (De Becker et al.\ \cite{DeB}). Most of these secondary sources were considered to be PMS stars or at least PMS candidates (Rauw et al.\ \cite{GR03}, hereafter Paper I). To ascertain the status of these objects, we obtained optical spectroscopy of a subsample of the X-ray selected objects as well as multi-band photometry of the cluster as a whole. The present paper reports the analysis of these data. 

\section{Observations \label{obs}}
Low-resolution spectra of a set of X-ray selected stars in NGC\,6383, taken from Paper I, were obtained on the night of June 22-23, 2004 with the EMMI instrument mounted on ESO's 3.5\,m New Technology Telescope (NTT) at La Silla. The EMMI instrument was used in the Red Imaging and Low Dispersion Spectroscopy (RILD) mode with grism \# 5 (600 grooves\,mm$^{-1}$), providing a wavelength coverage from about 3800 to 7020\,\AA. This instrumental configuration yields a spectral resolution of 5.0\,\AA\ ({\tt FWHM} of the He-Ar lines in the wavelength calibration exposures). The targets selected for spectroscopy were the brightest X-ray sources found to have a single optical counterpart within the 8\,arcsec correlation radius used in Paper I. Their $V$ magnitudes range from about 11.3 to 17.9. The exposure times were between 5 and 40\,min depending on the magnitude of the targets. The data were reduced in the standard way using the {\tt long} context of the {\sc midas} package. The observing conditions were photometric over most parts of the night and we thus performed a relative flux calibration of our spectra against an observation of the spectrophotometric standard star LTT\,9239 (Hamuy et al.\ \cite{Hamuy}).\\

Photometric observations of the cluster were obtained with the Wide Field Imager (WFI) instrument at the ESO/MPG 2.2\,m telescope at La Silla during two observing runs in September 2003 (service mode) and June 2004 (visitor mode). The WFI instrument has a field of view of about $34' \times 33'$ covered by a mosaic of $4 \times 2$ CCD chips with a pixel size of 0.238\,arcsec/pixel. The frames were taken through a set of $U\,B\,V\,(R\,I)_c\,H\alpha$ filters (see Table\,\ref{filtres}) with various exposure times (from 1 second to about 10 minutes, depending on the filter) to allow measurements of both faint and moderately bright objects. Our photometric data are complete to magnitudes of about 20.5, 22.5, 22.0, 21.5, 20.5, and 20.5 for the $U$, $B$, $V$, $R_c$, $I_c$, and H$\alpha$ filters, respectively. 

The data were bias subtracted and flat fielded using the {\sc iraf} package. The  raw images consist of dithered observations performed to fill the gaps between the WFI CCDs. The photometry in the natural system was obtained with the {\sc daophot} software in two steps: a first pass on stacked, registered images in all filters allowed us to produce a deep catalogue of sources. In a second pass, the photometry of these sources was obtained for the individual CCDs with apertures of 4, 5, 8, and 11 pixel diameter (i.e., 0.95, 1.2, 1.9, and 2.6\,arcsec). The larger aperture data were used to perform all-sky photometry. After a number of iterations, satisfactory zeropoints and extinction coefficients were obtained. To determine the latter, seven Stetson ($B\,V\,R\,I$) and Landolt ($B\,V\,R\,I$) standard fields were observed during the same nights. In addition, $U\,B\,V\,R\,I$ data for 8 stars in NGC\,6383 were obtained from FitzGerald et al.\ (\cite{Fitz}), Th\'e et al.\ (\cite{The}) and Zwintz et al.\ (\cite{Zwintz}). The data of the individual frames were then homogenized using bright isolated stars as secondary standards with values determined from the all-sky photometry. The astrometry was established by matching the instrumental coordinates with the 2MASS point source catalogue. The 1.2\,arcsec aperture is well suited to our analysis in terms of both accuracy and sensitivity to crowding in the field. In the following, we use the photometry corresponding to this aperture, except when stated otherwise.

\begin{table}
\caption{The filters used in the photometric campaigns with the WFI.\label{filtres}}
\begin{tabular}{c c}
\hline
Filter & ESO number \\
\hline
$U$ & BB\#U/50\_ESO877 \\
$B$ & BB\#B/123\_ESO878 \\
$V$ & BB\#V/89\_ESO843 \\
$R_c$ & BB\#Rc/162\_ESO844 \\
$I_c$ & BB\#I/203\_ESO879 \\
$H\alpha$ & NB\#Halpha/7\_ESO856 \\
\hline
\end{tabular}
\end{table}
\begin{figure}[htb]
\begin{center}
\resizebox{9cm}{!}{\includegraphics{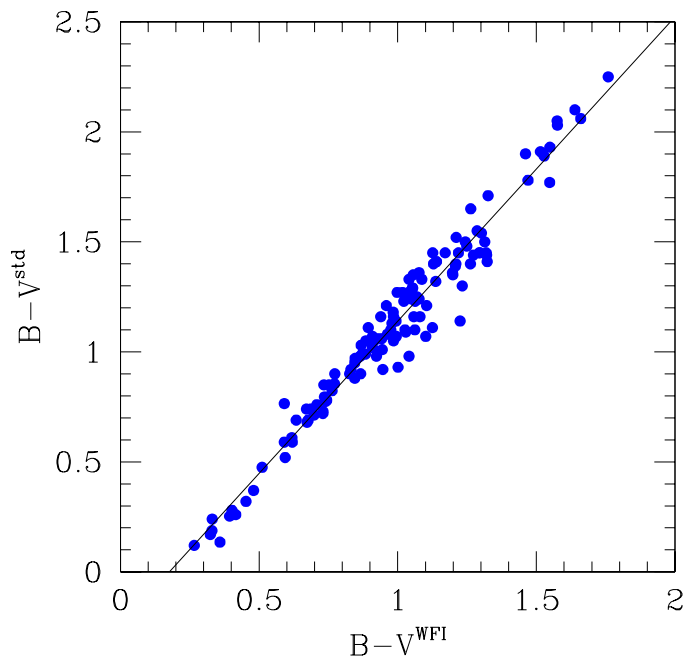}}
\end{center}
\caption{Comparison between the $B-V$ colour index in the WFI system and the standard system for 156 stars selected as the best observed among those in common with Paunzen et al.\ (\cite{Paunzen}). The solid line yields our best-fit colour transformation.\label{coltrans}}
\end{figure}

The passbands of the WFI filters, and in particular that of the $B$ filter, are rather non-standard. Unfortunately, the limited number of observed standard stars did not cover a sufficient range of spectral types to establish precise colour transformations from these standards. Therefore, we compared our WFI photometric data in the natural system with published Johnson-Cousins photometry of stars in common in NGC\,6383. In this context, we established a colour transformation (Fig.\,\ref{coltrans}) based on the comparison between our $V^{\rm WFI}$ magnitudes and $(B - V)^{\rm WFI}$ colours and the standard $B\,V$ photometry for 156 stars in common with the (online) Table\,1 of Paunzen et al.\ (\cite{Paunzen}). The resulting relations are\footnote{The zeropoints in these relations only apply to our reduction of this specific data set and cannot be compared to other reductions of other WFI data.} 
$$(B - V)^{\rm std} = -0.227 + 1.418\,(B - V)^{\rm WFI},$$ 
$$V^{\rm std} = V^{\rm WFI} + 0.053 - 0.101\,(B - V)^{\rm WFI}.$$ 
The coefficients of the colour terms of these relations are in reasonable agreement with those proposed by ESO on the WFI instrument website\footnote{http://www.eso.org/sci/facilities/lasilla//instruments/wfi/inst/zero\-points/ColorEquations}. Unfortunately, there are much fewer available data for the calibration of the $U$, $R_c$, and $I_c$ filters. We therefore adopted the coefficients of the colour terms following the transformations given by ESO and only determined the zeropoints by comparing the WFI with the existing standard colours of seven stars. The resulting relations are
$$(U - V)^{\rm std}   = 1.08\,(U - V)^{\rm WFI} + 0.002 + 0.02\,(B - V)^{\rm std},$$ 
$$(V - R_c)^{\rm std} = 0.98\,(V - R)^{\rm WFI} + 0.017 - 0.09\,(B - V)^{\rm std},$$ 
$$(V - I_c)^{\rm std} = 0.94\,(V - I)^{\rm WFI} + 0.023 - 0.08\,(B - V)^{\rm std}.$$ 
In the following, we applied these transformations to all of our data. Unless stated otherwise, all colours given in this paper refer to the transformed colours.\footnote{Table 2 with the standard photometric data of those sources that are detected in all filters is available at the CDS.}

\begin{figure*}[h!tb]
\begin{minipage}{7.5cm}
\begin{center}
\resizebox{!}{7cm}{\includegraphics{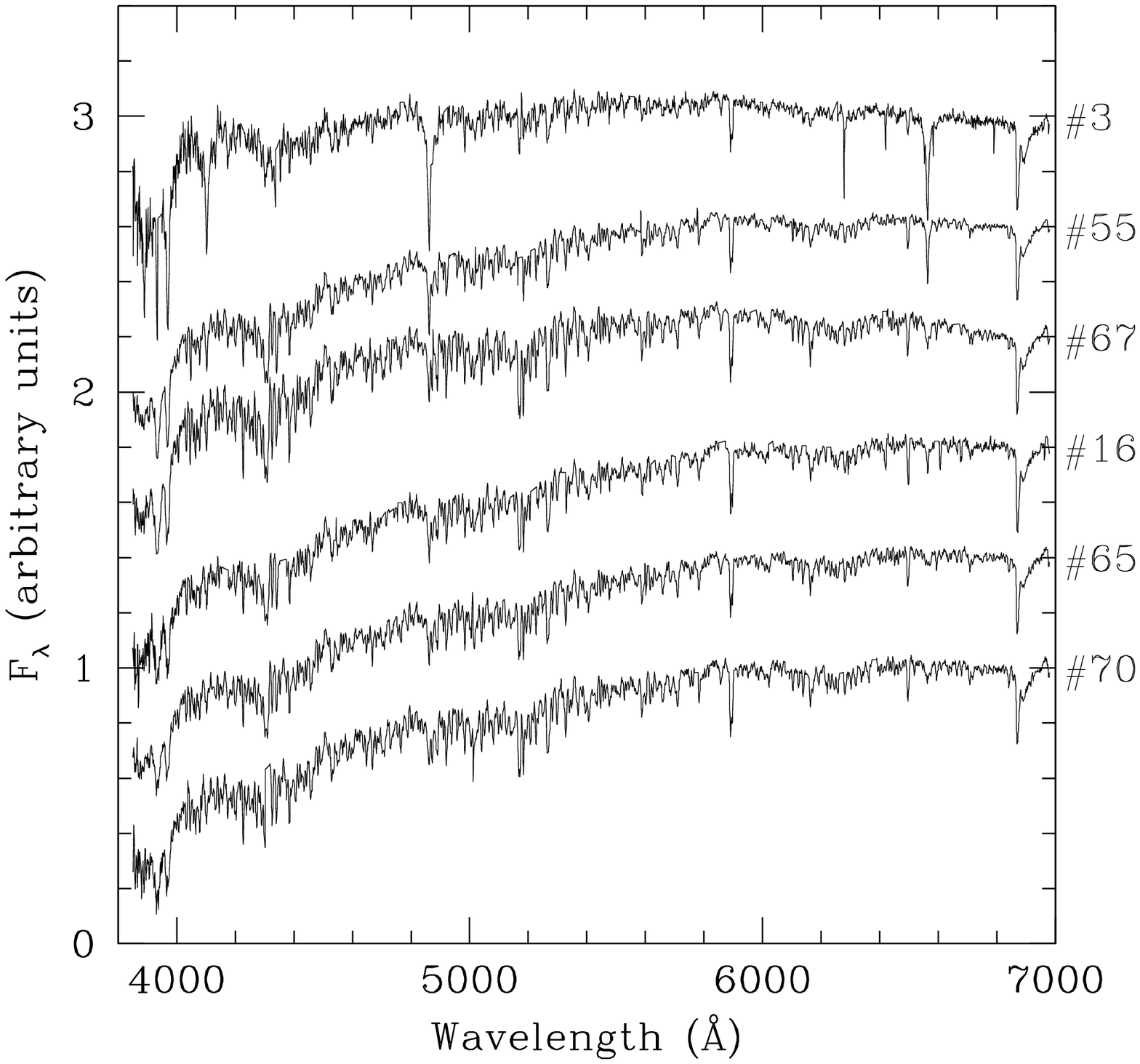}}
\end{center}
\end{minipage}
\begin{minipage}{7.5cm}
\begin{center}
\resizebox{!}{7cm}{\includegraphics{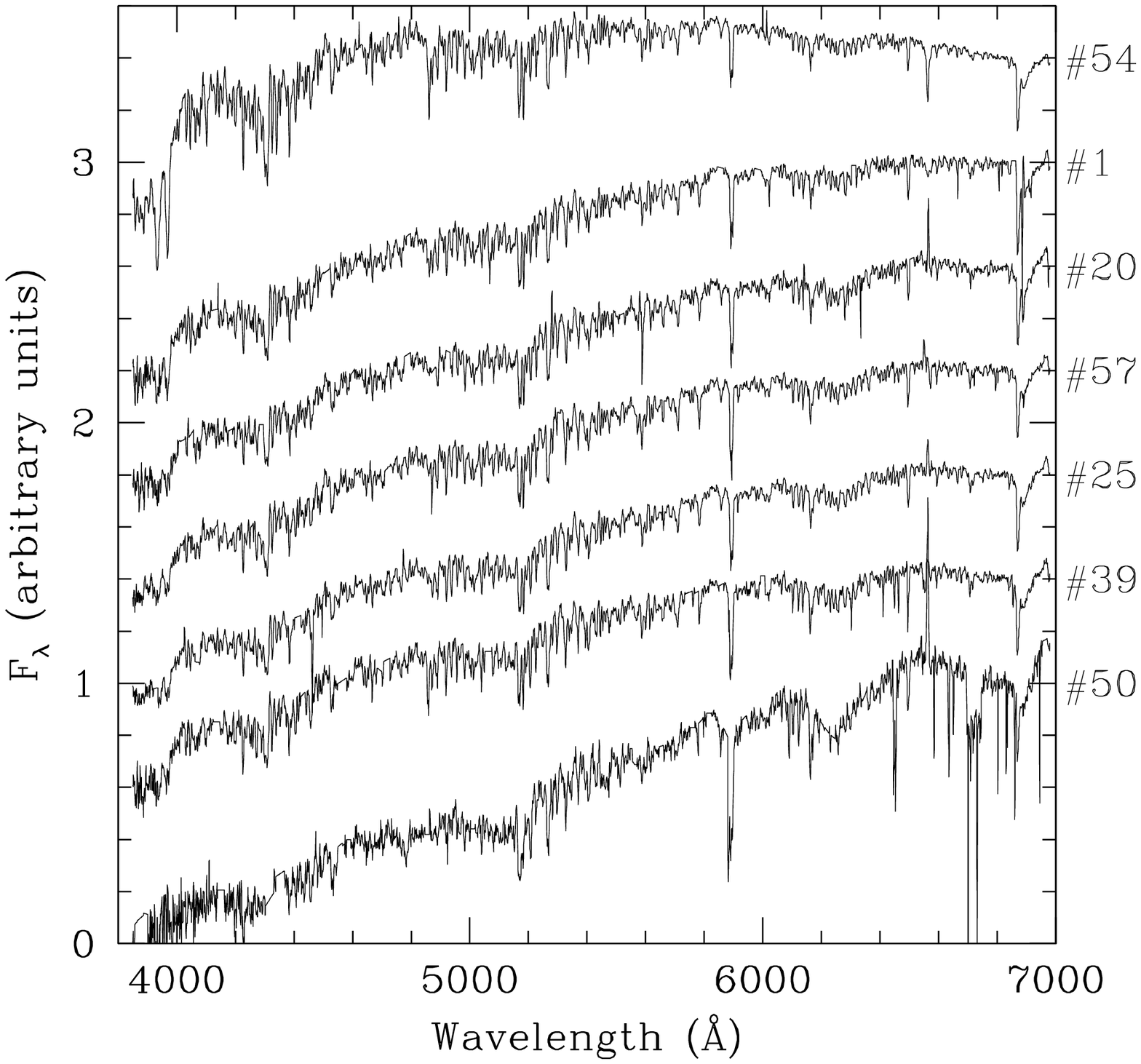}}
\end{center}
\end{minipage}
\begin{minipage}{3cm}
\begin{center}
\resizebox{!}{7cm}{\includegraphics{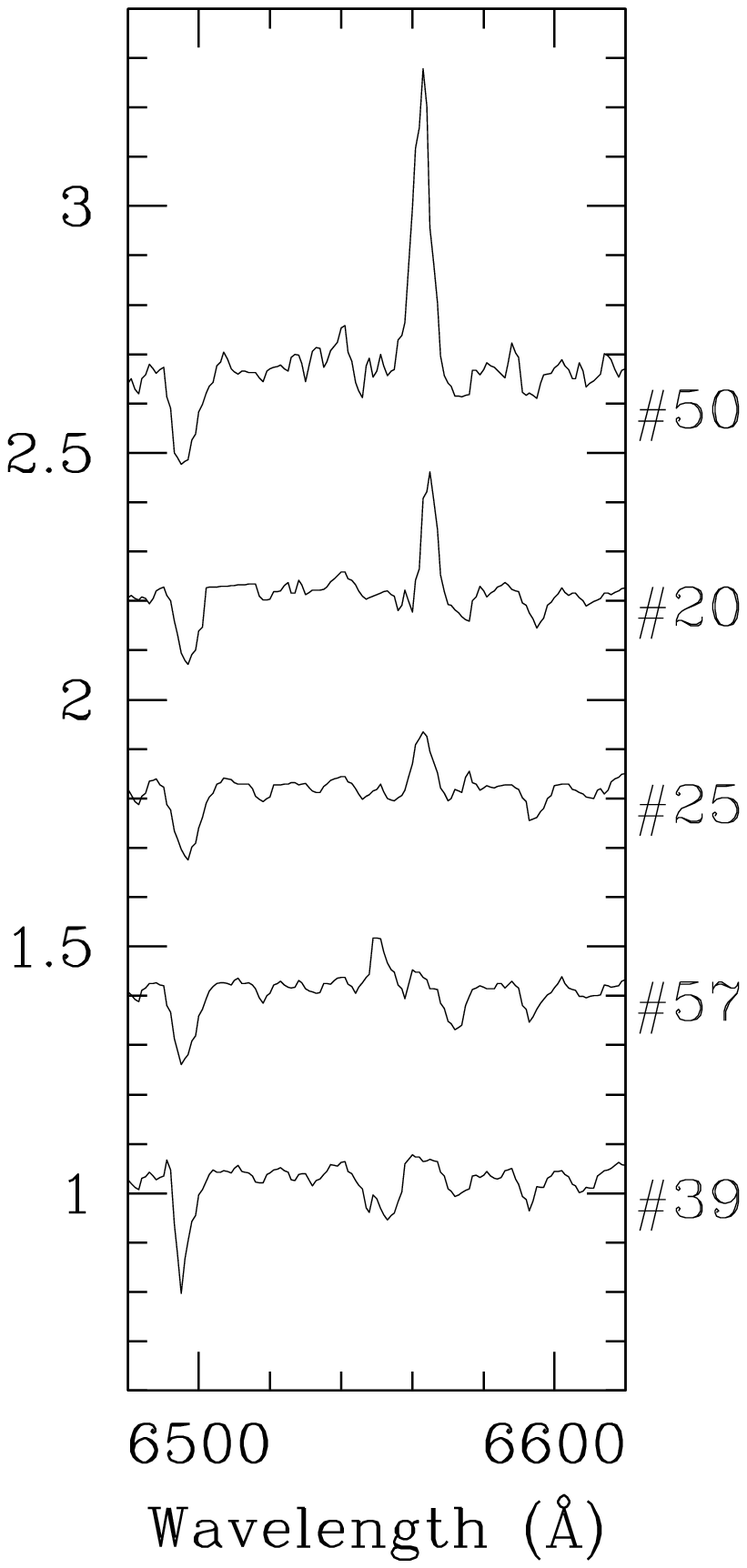}}
\end{center}
\end{minipage}
\caption{Left and middle panel: flux-calibrated spectra of the F-K type stars in our spectroscopic sample of NGC\,6383. The fluxes were arbitrarily normalized to unity at the wavelength of 6820\,\AA\ and the spectra were shifted vertically by 0.4 units for clarity. The spectral types become progressively later from left to right and from top to bottom. The rightmost panel shows a zoom on the spectral region around the H$\alpha$ line for those stars that are emitters or emission candidates.\label{fig-1}}
\end{figure*}

Because of its special passband definition and a lack of suitable standard fields, the H$\alpha$ filter in use at the WFI cannot be directly tied to any existing standard photometric system. In the natural system, the $R_c - H\alpha$ zeropoint was at first fixed arbitrarily to zero for the standard stars. In a second step, we re-calibrated the zeropoint by comparing the observed $R_c - H\alpha$ and dereddened $V - I_c$ indices of 9 stars whose spectra do not exhibit H$\alpha$ emission with the $R_c - H\alpha$ versus $V - I_c$ relation of emission-free main-sequence stars in NGC\,2264 as determined by Sung et al.\ (\cite{Sung}). To express the $R_c - H\alpha$ indices in the system of Sung et al.\ (\cite{Sung}), we need to subtract $(4.71 \pm 0.06)$ from the observed indices. All values of the $R_c - H\alpha$ index quoted in this paper and all plots involving this index refer to the Sung et al.\ (\cite{Sung}) system.

\section{Spectroscopy}
To establish the MK spectral types of the targets in NGC\,6383, we used the digital spectral classification atlas compiled by R.O.\ Gray available on the web\footnote{http://nedwww.ipac.caltech.edu/level/Gray/frames.html}. The details of the classification are given for each star in Appendix\,\ref{app}. The results are summarized in Table\,\ref{class}. 

The majority of the spectroscopically studied objects are late-type (F-K) main-sequence or slightly more luminous stars (see Fig.\,\ref{fig-1} and Table\,\ref{class}). The exceptions are the mid B-type star\footnote{Throughout this paper, we shall use the numbering convention introduced in Paper I to designate the X-ray sources in the cluster, except when stated otherwise.} \#56 and the early-type B star \#76 (see Fig.\,\ref{fig-2}). 

\subsection{Late-type stars}
For most stars of our spectroscopic sample, H$\alpha$ is detected in absorption. The exceptions are stars \#20, 25, 50, and possibly \#57 and 39. We checked the existence of these emission lines on the original two-dimensional long-slit spectra to avoid incorrectly identifying a cosmic ray hit or poorly corrected nebular H$\alpha$ emission. Although nebular emission exists around all targets of our sample, it is removed by sky subraction and the nebular line is too weak to account for the strength of the observed stellar emission features. We emphasize that the strongest emission is seen in the spectrum of the counterpart of X-ray source \#50. This object experienced an X-ray flare during the {\it XMM-Newton} observation (see Paper I). The situations for stars \#57 and \#39 are not entirely clear. In the former, we observe two weak emission peaks bluewards of the rest-frame wavelength of the H$\alpha$ line with heliocentric velocities of $-564$ and $-71$\,km\,s$^{-1}$. Although our inspection of the 2D-spectra confirmed that these features are not artefacts, the detection of this blueshifted emission structure needs to be confirmed by additional observations. Finally, in the spectrum of star \#39, a broad absorption feature is observed blueward of the H$\alpha$ rest wavelength, while the line itself seems to be filled in by emission.  
 
All spectroscopically observed objects with H$\alpha$ emission in their spectrum are of spectral type K (see Fig.\,\ref{fig-1}), while all stars of earlier spectral type display the H$\alpha$ line in absorption. We also note that the equivalent widths (EWs) of these emission lines are rather modest. For instance, they are much weaker than the 10\,\AA\ equivalent width conventionally adopted as the limit between the objects of the wTTS and cTTs categories (e.g., Preibisch et al.\ \cite{Preibisch}). Therefore, these objects are candidate wTTs.  

Although the spectral resolution of our data is insufficient to resolve the possible Li\,{\sc i} $\lambda$\,6708 line from the neighbouring Fe\,{\sc i} lines, we note that the spectra of the majority of the targets later than spectral type F (especially stars \#1, 16, 25, 39, 55, 57, 65, 67, and 70) display a rather strong absorption feature at the wavelength of the Li\,{\sc i} line. The unresolved blend with the Fe\,{\sc i} lines probably accounts for the asymmetric shape of the observed profiles. The possible detection of Li in the spectra of the late-type stars of our sample is compatible with a PMS status. We note that all spectra (except \#56 and \#76) display a significant absorption feature at about 6104\,\AA. This feature is unlikely to be associated with the Li line observed in Population II stars, whose intensity should be very weak (see e.g., Ford et al.\ \cite{Ford}). This absorption could instead be caused by a blend of Ca\,{\sc i} $\lambda$\,6103 with Fe\,{\sc i}.
   
Finally, we emphasize that star \#55 (= FJL\,23) was previously suggested to be a foreground object (Th\'e et al.\, \cite{The}; Rauw et al.\ \cite{GR03}), but this conclusion is not confirmed by our present analysis. The photometric and spectroscopic properties of this star are indeed consistent with it being a member of NGC\,6383.
  
\subsection{Early-type stars}
The OB star \#76 appears heavily absorbed and must therefore belong to a population of stars located behind NGC\,6383. Our photometric measurements indeed yield $E(B - V) \simeq 2.5$ for this star. The NGC\,6383 cluster is understood to be located in front of a dust cloud (Lloyd Evans \cite{LE}) and star \#76 is therefore likely a background object located either inside or behind this cloud. 
\begin{figure}[htb]
\begin{center}
\resizebox{9cm}{!}{\includegraphics{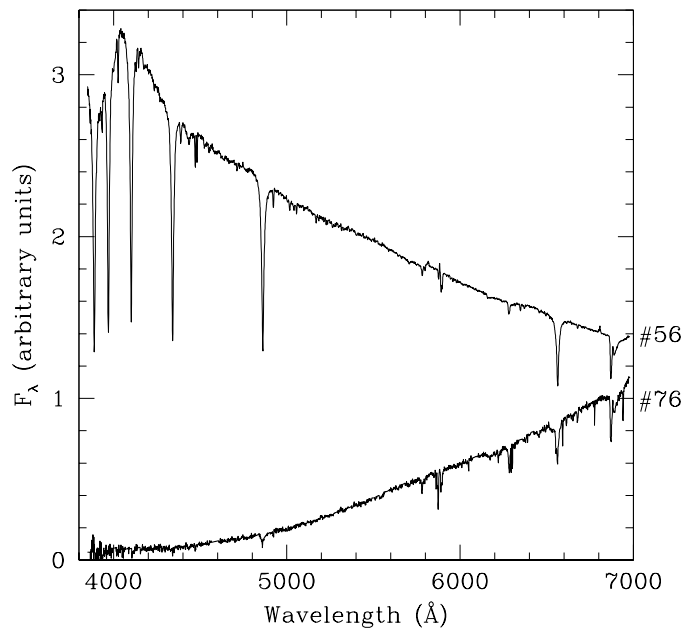}}
\end{center}
\caption{Flux-calibrated spectra of the two early-type stars in our spectroscopic sample of NGC\,6383. Star \#76 probably has an earlier spectral type than \#56, but is much more absorbed. The fluxes are presented in the same way as in Fig.\,\ref{fig-1}.\label{fig-2}}
\end{figure}

The case of star \#56 (= FJL\,24, FitzGerald et al.\ \cite{Fitz}) is also quite interesting. Mid to late-type B stars are usually not strong X-ray emitters. Using the bolometric correction for a B5-7\,V star and the X-ray flux from Paper I, we derive a $\log{L_{\rm X}/L_{\rm bol}}$ value of $-5.13 \pm 0.11$ for this star. To evaluate the X-ray luminosity, we converted the EPIC-pn count rate given in Paper I into an unabsorbed flux of $(8.2 \pm 1.9) \times 10^{-14}$\,erg\,cm$^{-2}$\,s$^{-1}$ in the 0.5 -- 10\,keV band, assuming a thermal plasma at a temperature of kT = 0.5\,keV and an absorption by an interstellar neutral hydrogen column of $1.91 \times 10^{21}$\,cm$^{-2}$. We note that our estimate of the X-ray flux would change by less than 20\% regardless of the temperature of the emitting plasma in the range 0.3 -- 2\,keV. Assuming a distance of 1.3\,kpc, the resulting X-ray luminosity amounts to $(1.7 \pm 0.4) \times 10^{31}$\,erg\,s$^{-1}$, which places this object clearly among the X-ray bright late B-type stars (see e.g., Bergh\"ofer et al.\ \cite{BSDC}, Sana et al.\ \cite{lxlbol}, and Naz\'e \cite{YN}). We note that this value of the X-ray luminosity would also be rather high if the emission was originating in an unresolved PMS companion. 

\addtocounter{table}{1}
\begin{table*}
\caption{Summary of the properties of the targets of our spectroscopic sample. The last column yields the cross identification with objects previously studied by FitzGerald et al.\,(\cite{Fitz}) as well as some information about objects with H$\alpha$ emission.\label{class}}
\begin{tabular}{c c c c c c c c}
\hline
Star & Spectral Type & H$\alpha$ & $V$ & $B-V$ & $V-I_c$ & $R_c-H\alpha$ & Notes \\
\hline
\#1 & K0\,V & abs. & 15.04 & 1.16 & 1.45 & $-4.70$ & \\
\#3 & F5\,V & abs. & 13.51 & 0.77 & 1.07 & $-4.77$ & \\
\#16 & G0--5\,V--IV & abs. & 14.96 & 1.10 & 1.37 & $-4.72$ & \\
\#20 & K0\,V & em. & 15.84 & 1.25 & 1.67 & $-4.72$ & $EW = -1.0$\,\AA,$v = 85$\,km\,s$^{-1}$ \\
\#25 & K2\,V & em. & 15.09 & 1.25 & 1.62 & $-4.54$ & $EW = -0.5$\,\AA,$v = 5$\,km\,s$^{-1}$ \\
\#39 & K2--4\,V & P-Cyg? & 15.15 & 1.25 & 1.54 & $-4.70$ & = FJL\,9 \\
\#50 & K7\,V & em. & 17.88 & 1.61 & 2.09 & $-4.65$ & $EW = -3.0$\,\AA,$v = -14$\,km\,s$^{-1}$ \\
\#54 & G7\,V & abs. & 12.77 & 0.85 & 1.00 & $-4.71$ & \\
\#55 & G0--1\,IV & abs. & 13.81 & 0.97 & 1.20 & $-4.72$ & = FJL\,23 \\
\#56 & B5--7\,V & abs. & 11.35 & 0.26 & 0.40 & $-4.85$ & = FJL\,24 \\
\#57 & K0--5\,III & double peak em.? & 15.49 & 1.25 & 1.56 & $-4.72$ & $EW = -0.3$\,\AA,$v = -564$, $-71$\,km\,s$^{-1}$ \\
\#65 & G5\,IV--III & abs. & 14.02 & 1.05 & 1.30 & $-4.72$ & \\
\#67 & G5--K0\,V & abs. & 14.21 & 0.91 & 1.12 & $-4.65$ & \\
\#70 & G5--K0\,IV--III & abs. & 15.19 & 1.08 & 1.33 & $-4.75$ \\
\#76 & O9--B5 & abs. & 17.81 & 2.21 & 2.89 &  & $E(B-V) \simeq 2.5$ \\
\hline
\end{tabular}
\end{table*}  

\section{Photometry}
Photometric data for a total of about 80\,000 sources were extracted. We have cross-correlated these data with the 2MASS near-IR catalogue (Skrutskie et al.\,\cite{Skrutskie}) and found about 18\,000 matches within 0.5\,arcsec, 7000 of which have good quality near-IR photometry. Among the latter, 2639 objects have reliable WFI photometry in all filters including the H$\alpha$ filter.

\begin{figure*}[htb]
\begin{minipage}{9cm}
\begin{center}
\resizebox{9cm}{!}{\includegraphics{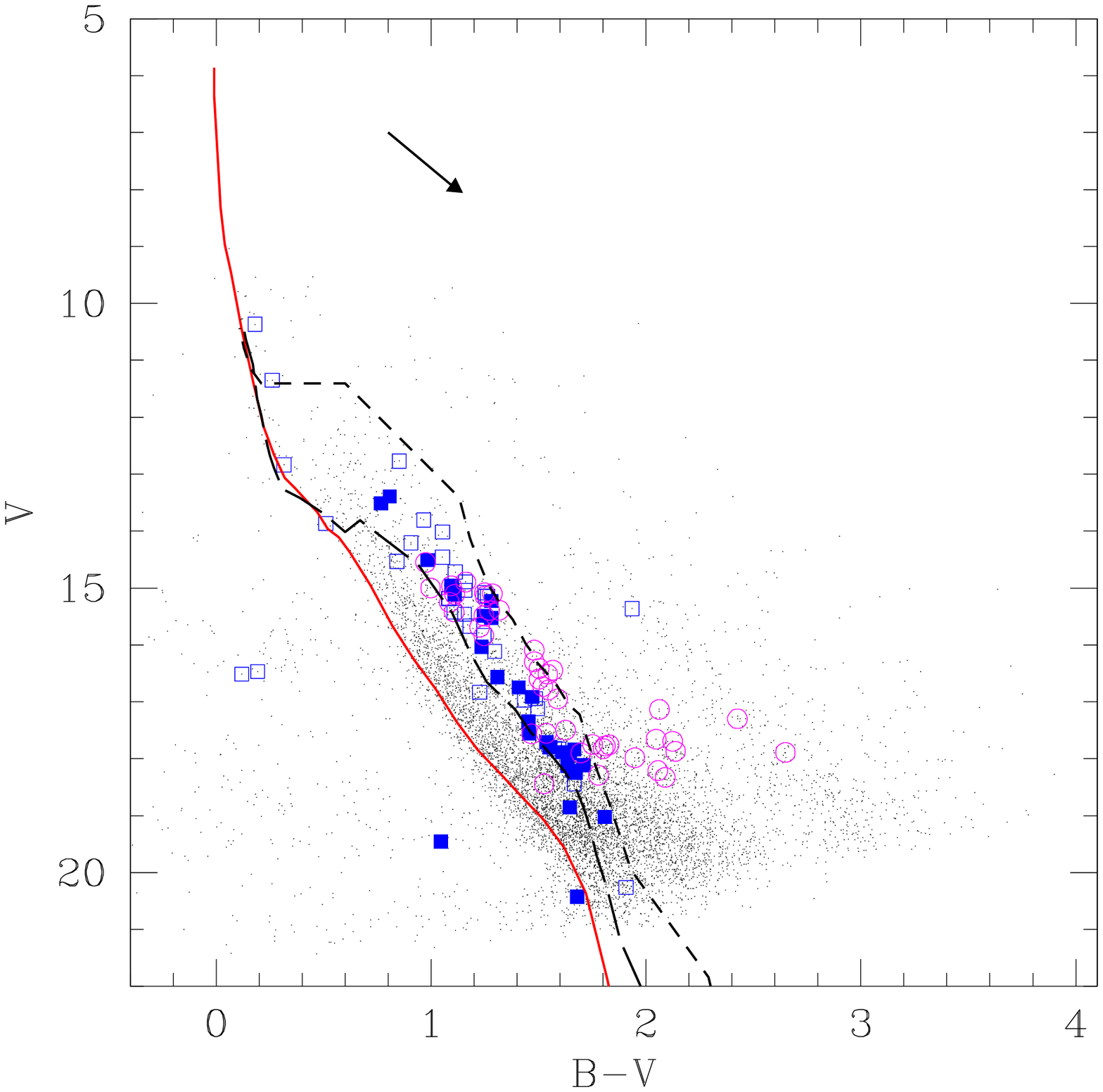}}
\end{center}
\end{minipage}
\hfill
\begin{minipage}{9cm}
\begin{center}
\resizebox{9cm}{!}{\includegraphics{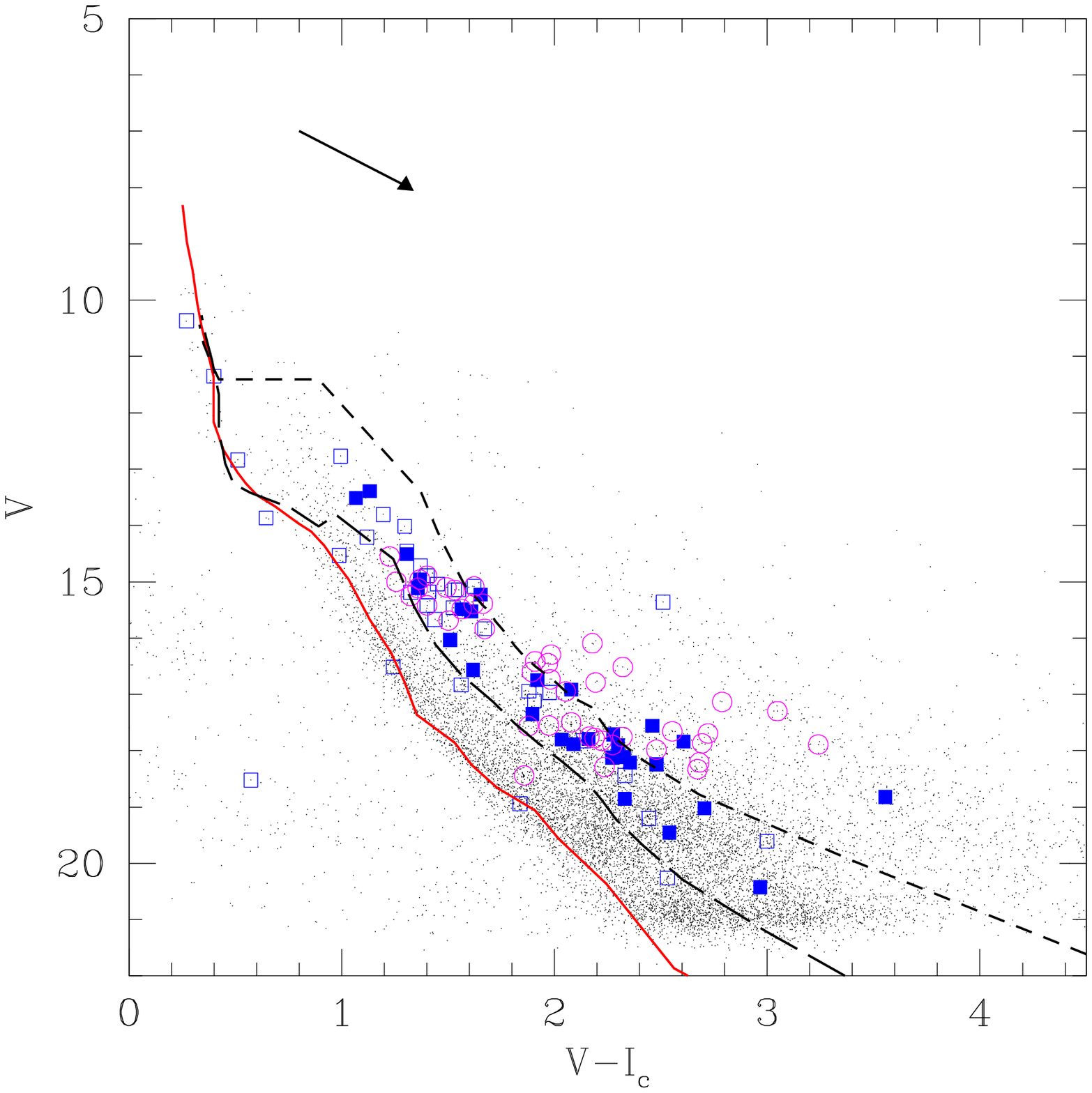}}
\end{center}
\end{minipage}
\caption{Colour-magnitude diagrams of the WFI photometric data with errors of less than 0.15 on $V$ and less than 0.21 in the relevant colour. Single and multiple counterparts of the X-ray sources from Paper I are shown as filled and open squares, respectively, while the PMS candidates from Paunzen et al.\ (\cite{Paunzen}) are shown as open circles. The main-sequence relation was shifted to account for a distance modulus of 10.57 and a reddening of $E(B - V) = 0.32$. The reddening vectors are shown for each diagram and the short and long-dashed lines yield the 1.5 and 10\,Myr isochrones, respectively, from Siess et al.\ (\cite{Siess}).\label{figBV}}
\end{figure*}

Cluster members fainter than $V \sim 11$ are expected to be pre-main-sequence objects. Unfortunately, these objects populate a region of the colour-magnitude diagrams that is potentially highly contaminated by field stars unrelated to NGC\,6383 (see Fig.\,\ref{figBV}). Since the cluster is probably located in front of a dust cloud, we expect the background objects to appear redder than the true cluster members. 

\subsection{Cross-correlation with the X-ray catalogue}
We cross-correlated the catalogue of X-ray sources from Paper I with the list of sources detected in our photometric data. To derive the optimal correlation radius, we applied the technique of Jeffries et al.\ (\cite{Jeff97}). In this approach, the distribution of the cumulative number of catalogued sources as a function of the cross-correlation radius $r_X$ is given by
\begin{eqnarray*}
\Phi(d \leq r_X) & = & A\,\left[1 - \exp{\left(\frac{-r_X^2}{2\,\sigma^2}\right)}\right] \\
& + & (N - A)\,\left[1 - \exp(-\pi\,B\,r_X^2)\right],
\end{eqnarray*}
where $N$, $A$, $\sigma$, and $B$ indicate, respectively, the total number of
cross-correlated X-ray sources ($N = 76$), the number of true correlations,
the uncertainty in the X-ray source position, and the surface density of the catalogue of photometric sources. The best-fit model parameters are $A = 57.8$, $\sigma = 1.8$\,arcsec, and $B = 1.22 \times 10^{-2}$\,arcsec$^{-2}$. The latter value (independently determined from the best fit to the cross-correlation process) is in reasonable agreement with the true mean surface density of our photometric catalogue ($1.8 \times 10^{-2}$\,arcsec$^{-2}$).

The optimal correlation radius that includes the majority of the true correlations whilst simultaneously limiting contamination by spurious coincidences to about 15\%, is found to be about 5\,arcsec. The following X-ray sources have no optical counterpart in our photometric data: \#4, 5, 18, 19, 21, 26, 28, 40, 71, and \#72. Some of the sources (e.g.\ \#14 and \#36) which had no counterpart in the USNO catalogue (see Paper I), now have a faint optical counterpart in our data. We note that HD\,159176 (X-ray source \#44) is too bright to be reliably measured in our photometric data and was thus not included in the correlation process.

The colour-magnitude diagrams shown in Fig.\,\ref{figBV} indeed confirm that the bulk of the X-ray sources with an optical counterpart populate a locus where PMS stars are expected. We note that these diagrams also include the 44 PMS candidates identified by Paunzen et al.\ (\cite{Paunzen}) on the grounds of photometric criteria.   

\begin{figure*}[thb]
\begin{minipage}{6.2cm}
\begin{center}
\resizebox{6.2cm}{!}{\includegraphics{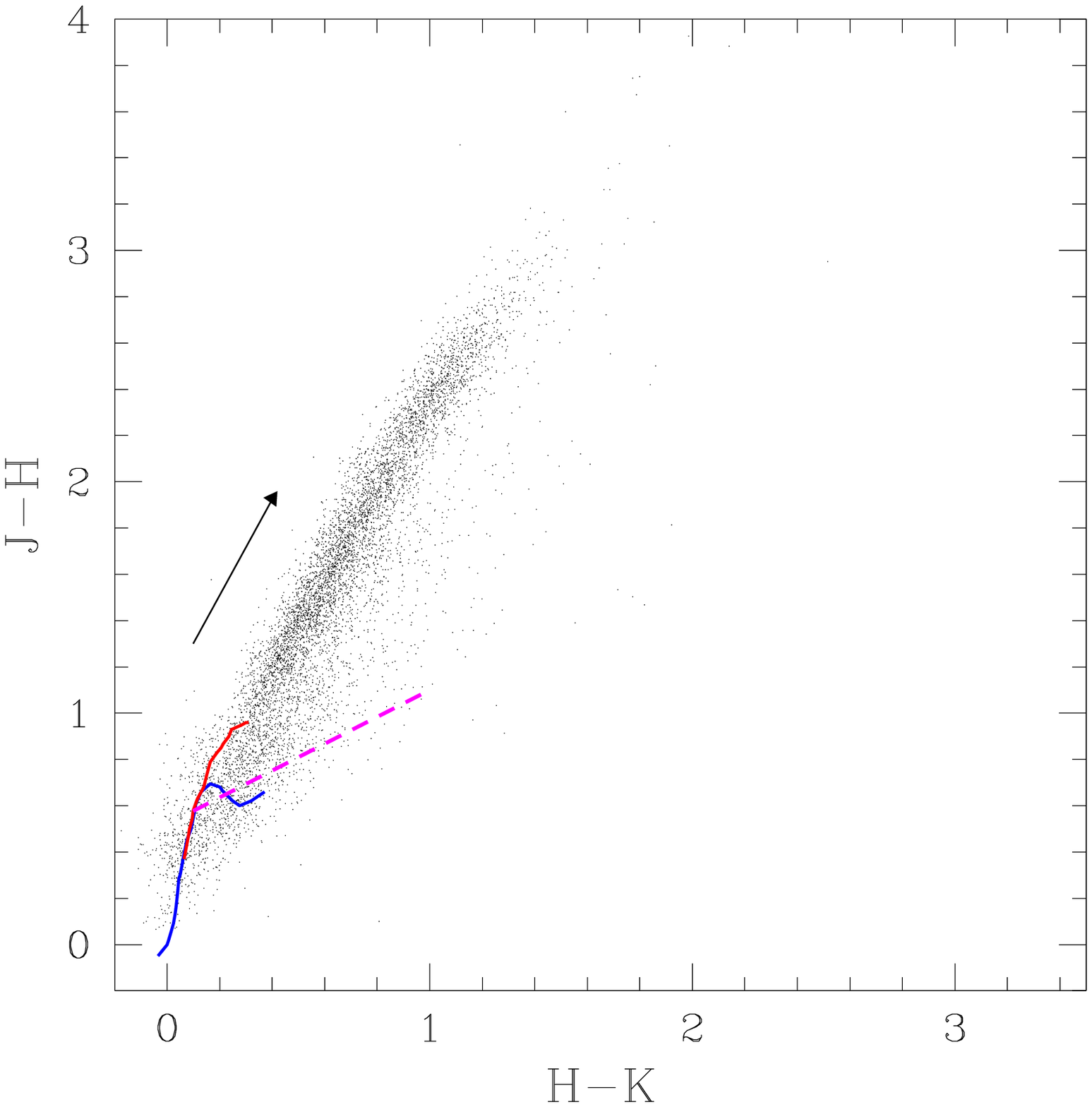}}
\end{center}
\end{minipage}
\begin{minipage}{6.2cm}
\begin{center}
\resizebox{6.2cm}{!}{\includegraphics{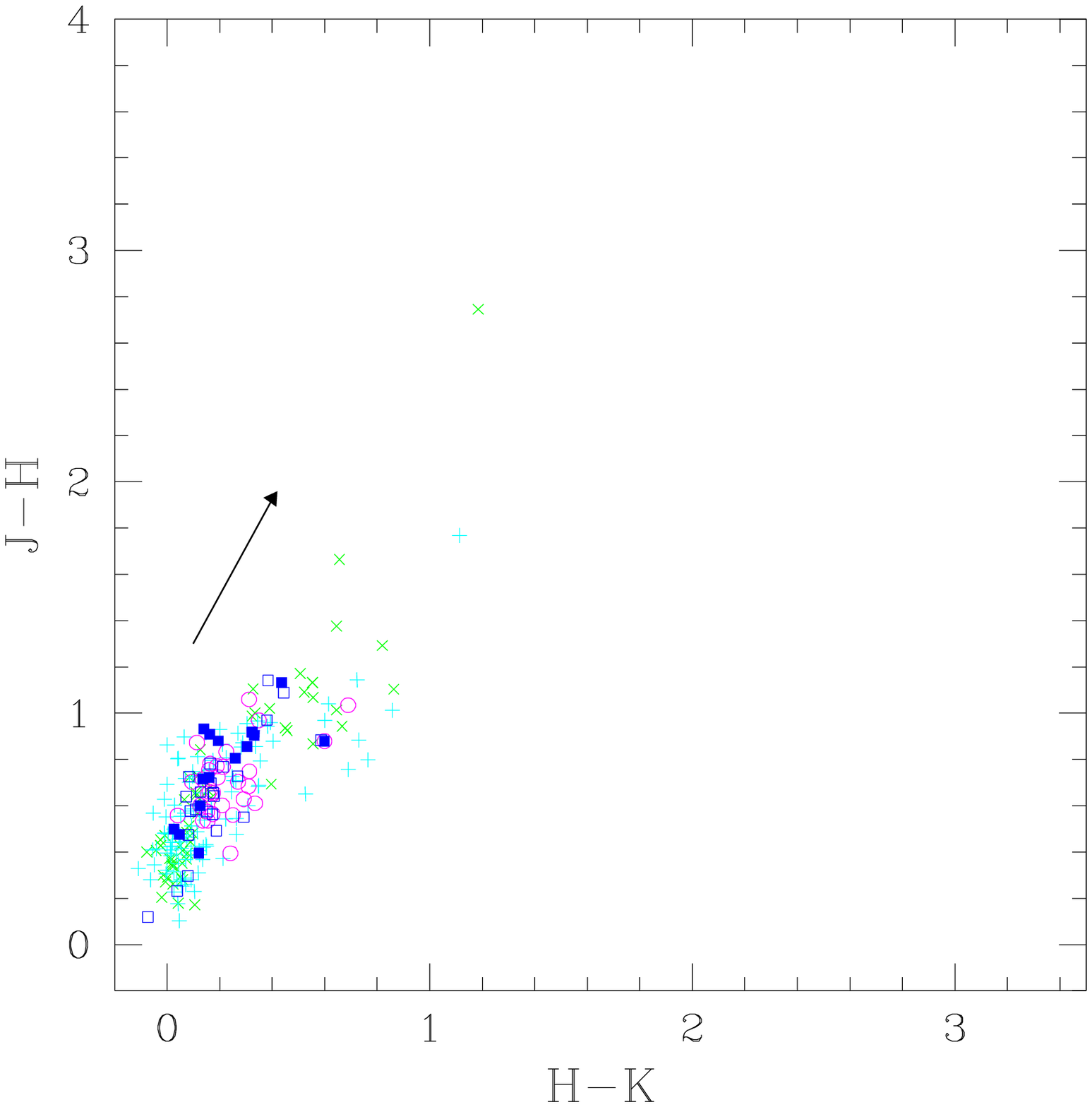}}
\end{center}
\end{minipage}
\begin{minipage}{6.2cm}
\begin{center}
\resizebox{6.2cm}{!}{\includegraphics{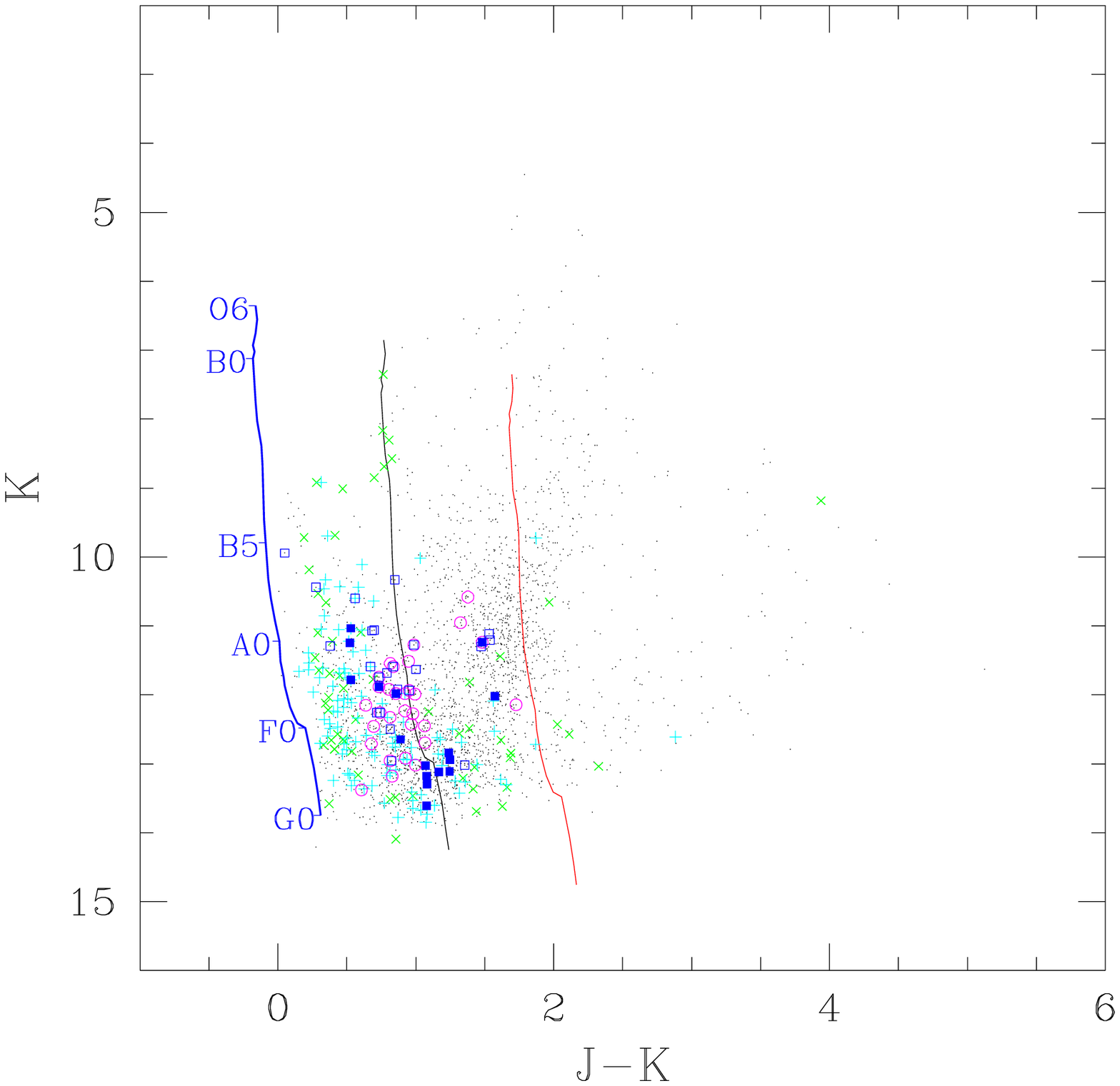}}
\end{center}
\end{minipage}
\caption{Left panel: colour-colour diagram of the 2MASS counterparts of the WFI objects. Only those sources with reliable near-IR photometry are shown. The solid lines yield the locus of the main-sequence and of the giant branch according to Bessell \& Brett (\cite{BB}), while the dashed straight line indicates the locus of the unreddened cTTs following Meyer et al.\ (\cite{Meyer}). The reddening vector is shown for $A_{Ks} = 0.5$. Middle panel: Location of the X-ray sources from Paper I (filled and open squares), PMS candidates from Paunzen et al.\ (\cite{Paunzen}, open circles) and H$\alpha$ emitters ($\times$ symbols) and candidates ($+$ symbols) in the $J\,H\,K$ colour-colour diagram. The other symbols have the same meaning as in Fig.\,\ref{figBV}. Right panel: colour-magnitude diagram. The main-sequence is shown for a distance modulus of 10.57 and for three different values of the reddening (from left to right, $A_{Ks} = 0.0$, 0.5, and 1.0). \label{2mass}}
\end{figure*}

\subsection{Near infrared photometry}
Based on the cross-correlation with the 2MASS data, we can compile a near-IR colour-colour diagram. For this purpose, we applied the colour transformations of Carpenter (\cite{Carpenter}, updated in 2003\footnote{http://www.ipac.caltech.edu/2mass/index.html}) to convert the 2MASS colours into the homogenized $J\,H\,K$ photometric system of Bessell \& Brett (\cite{BB}). The results are illustrated in Fig.\,\ref{2mass}. In the left panel of this figure, we show the $J\,H\,K$ colour-colour diagram. The heavy solid lines yield the intrinsic colours of main-sequence stars and giants according to Bessell \& Brett (\cite{BB}), while the dashed straight line yields the locus of dereddened colours of classical T\,Tauri stars according to Meyer et al.\ (\cite{Meyer}). 

With $l = 355.69^{\circ}$ and $b = +0.04^{\circ}$, the NGC\,6383 field lies just outside the area investigated by Nishiyama et al.\ (\cite{Nishiyama}) in their study of the extinction law towards the Galactic centre. Nevertheless, we assume here that their results also apply to our targets. In doing so, we find that the apparent reddening of the objects in the field of NGC\,6383 spans a very wide range: about 2.4 in $A_{Ks}$, which corresponds roughly to 20 magnitudes in $A_V$. It is most interesting that there appears to be a continuous distribution of reddening values. We already pointed out that the cluster is likely located at the front side of a large molecular cloud. The results in Fig.\,\ref{2mass} imply that we observe a population of sources from different depths inside this molecular cloud. This situation compromises any attempt to define a straightforward cluster membership criterion based on the reddening.

In the middle panel of Fig.\,\ref{2mass}, we have also plotted the near-IR counterparts of those stars that are either selected as X-ray sources, PMS candidates from Paunzen et al.\,(\cite{Paunzen}), or H$\alpha$ emitters (see next subsection). This figure shows that the vast majority of the X-ray sources are subject to a rather moderate reddening. At the same time, we find no obvious correlation between H$\alpha$ emission and the existence of a near-IR emission: the majority of the possible H$\alpha$ emitters (see Sect.\,\ref{Ha}) have rather normal near-IR colours. 

Finally, the rightmost panel of Fig.\,\ref{2mass} displays the colour-magnitude diagram, where the main-sequence relation has been shifted for a distance modulus of 10.57. Again, the existence of a wide, continuous range of interstellar reddening is quite obvious. We note that the reddening of the bright cluster members ($E(B - V) = 0.32$) corresponds roughly to $A_{Ks} = 0.11$. 

\subsection{H$\alpha$ emission candidates identified from photometry \label{Ha}}
According to our current understanding of classical T\,Tauri stars, these objects are surrounded by a circumstellar disk that is truncated at its inner edge by the stellar magnetosphere. Most of the time, cTTs accrete at a rate well below $10^{-6}$\,M$_{\odot}$\,yr$^{-1}$ and their H$\alpha$ emission originates in accretion streams that are controlled by the magnetic field (see e.g., Schulz \cite{Schulz}).

We used the $R_c - H\alpha$ index, calibrated against the Sung et al.\ (\cite{Sung}) $R_c - H\alpha$ versus $V - I_c$ main-sequence relation (see Sect.\,\ref{obs}), to identify potential H$\alpha$ emitters. The relevance of the Sung et al.\ (\cite{Sung}) relation for our dataset was checked using synthetic photometry, obtained by numerical integration of the spectrophotometric data of main-sequence stars from Jacoby et al.\ (\cite{Jacoby}) through the WFI $V$ and $H\alpha$ filter passbands. The operation was performed for stars of spectral types from B4\,V to M5\,V. Our synthetic photometry is in excellent agreement with the Sung et al.\ main-sequence relation. We then repeated the integration by adding an H$\alpha$ emission line of increasing equivalent width to the Jacoby et al.\ (\cite{Jacoby}) spectra. In this way, we found that an H$\alpha$ equivalent width of 10\,\AA, which is usually adopted as the threshold for cTTs, corresponds to an $R_c - H\alpha$ index $0.24 \pm 0.04$ above the main-sequence relation. 

We therefore consider a star as an H$\alpha$ emission candidate if its $R_c - H\alpha$ colour index is between 0.12 and 0.24 above the main-sequence relation (corresponding roughly to an EW between 6 and 10\,\AA) and as an H$\alpha$ emitter if the corresponding colour index is more than 0.24 above the main-sequence relation. In this way, we arrive at a list of 924 H$\alpha$ candidates and 592 emitters. If we restrict ourselves to stars with photometric errors of less than 0.15\,mag in $V$ and less than 0.21 in $R_c - H\alpha$, these numbers become 475 and 189 for the H$\alpha$ candidates and emitters, respectively. The 0.06\,mag uncertainty in the zeropoint of the $R_c - H\alpha$ index translates into an uncertainty of a factor 1.8 in the numbers of emitters and candidates.

Since the objects in our data appear to have a range of extinction values, we checked that our selection criterion for H$\alpha$ emission is not biased by false detections which could be caused by a heavy reddening that produces an artificially high $R_c - H\alpha$ index. To do so, we evaluated the $\frac{(R_c + I_c)}{2}-H\alpha$ colour index, which has a different sensitivity to reddening than our $R_c - H\alpha$ criterion. The vast majority of objects selected with the Sung et al.\ (\cite{Sung}) criterion are also identified as candidates or emitters according to the $\frac{(R_c + I_c)}{2}-H\alpha$ colour.
\begin{figure*}[htb]
\begin{minipage}{9cm}
\begin{center}
\resizebox{9cm}{!}{\includegraphics{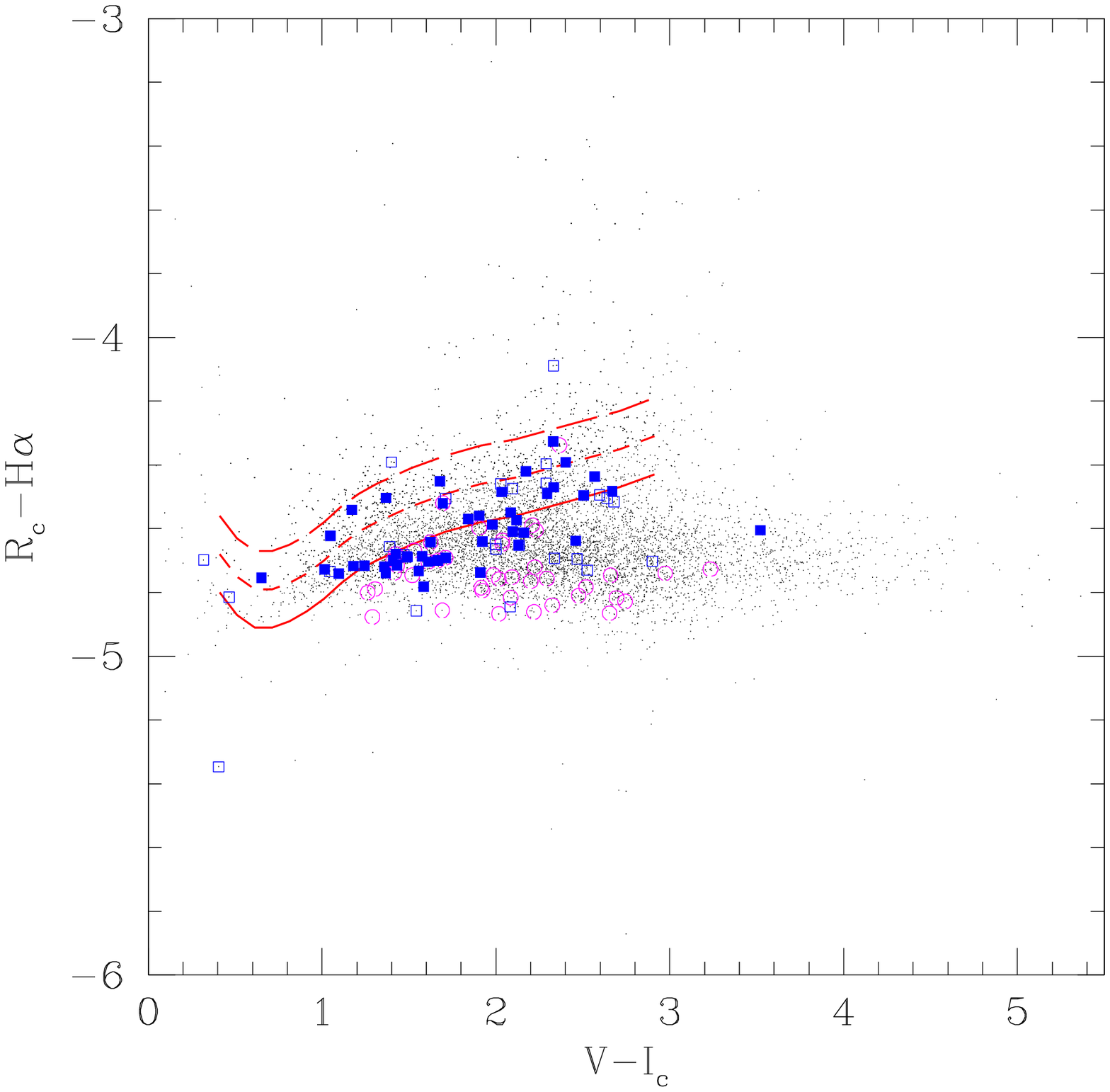}}
\end{center}
\end{minipage}
\hfill
\begin{minipage}{9cm}
\begin{center}
\resizebox{9cm}{!}{\includegraphics{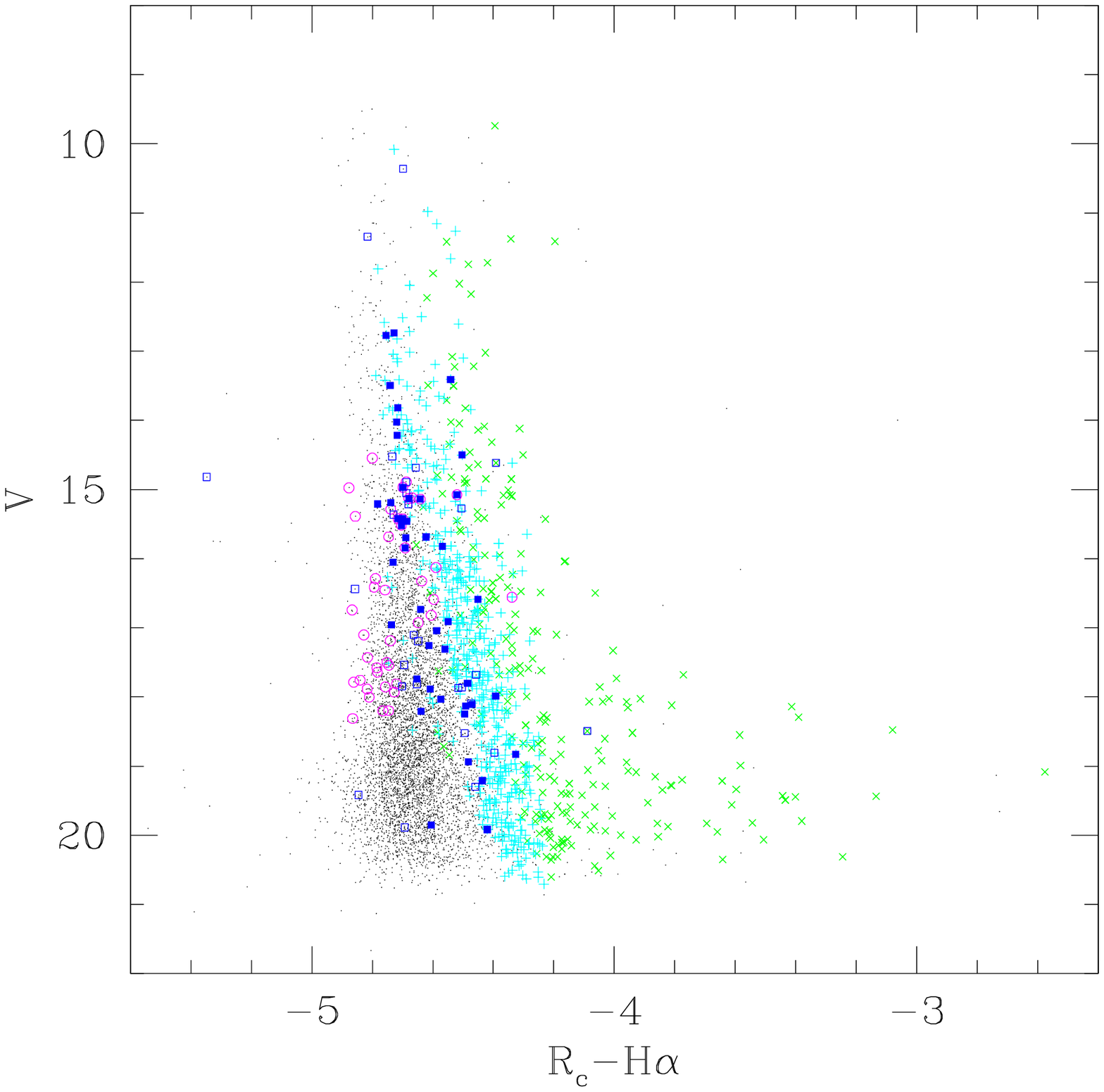}}
\end{center}
\end{minipage}
\caption{Left: the $R_c-H\alpha$ index is shown as a function of the $V - I_c$ colour. The symbols have the same meaning as in Fig.\,\ref{figBV}. The solid line indicates the relation for main-sequence stars as taken from Sung et al.\ (\cite{Sung}) and reddened by $E(V-I_c) = 0.512$. The short- and long-dashed lines yield the thresholds for H$\alpha$ candidates and emitters, respectively. Right: $V$ versus $R_c-H\alpha$ colour-magnitude diagram. The $+$ and $\times$ symbols represent H$\alpha$ emitter candidates and emitters, respectivley.\label{figVIRHa}}
\end{figure*}

From Fig.\,\ref{figVIRHa}, we see that neither the Paunzen et al.\ (\cite{Paunzen}) PMS candidates nor the X-ray sources display strong H$\alpha$ emission: there are only half a dozen X-ray sources in the H$\alpha$ emitter region. We note that those objects for which our spectroscopy revealed (rather weak) H$\alpha$ emission are not identified as emitters in our photometric data. This suggests that wTTs will be difficult to identify with narrow-band photometry (see also Dahm \& Simon \cite{DS} for a comparison of the efficiency of narrow-band photometry and slitless grism spectroscopy to identify weak H$\alpha$ emitters).

Whilst many emission candidates and emitters are rather faint objects, we see that H$\alpha$ emission is apparently not restricted to only the faintest (and hence least massive) stars. A few objects as bright as $V \sim 10$ -- $11$ are identified as H$\alpha$ emitters. The brightest is the Be star HDE\,317861 (B3 - 5\,Vne, Lloyd Evans \cite{LE}) with $V = 9.76$ and $R - H\alpha = -4.32$. Whilst the Herbig Ae star FJL\,4 (= V\,486 Sco, A5\,IIIp, Th\'e et al.\ \cite{The}) is classified as a candidate H$\alpha$ emitter with our criteria, indications of emission are found in neither our photometric nor spectroscopic data of FJL\,24 (= X-ray source \#56, B5-7). We note that the latter object was classified as B8\,Vne by FitzGerald et al.\ (\cite{Fitz}), since one observation out of six analysed by these authors displayed H$\beta$ emission. Our spectrum of this star exhibits strong Balmer absorption lines (see Fig.\,\ref{fig-2}).

The presence of H$\alpha$ emission does not automatically imply that a star is a member of the cluster. For instance, our sample of H$\alpha$ emitters and candidates may contain foreground, magnetically active, dMe stars, unrelated to the cluster. The incidence of dMe stars (with an H$\alpha$ EW $>$ 1\,\AA) increases towards lower masses and hence later spectral types (see Hawley et al.\ \cite{Hawley} and references therein): the observed frequency is about 10\% at spectral types earlier than M2, but increases towards about 60\% around spectral type M5. Furthermore, the dMe stars of the earlier spectral types appear on average 0.5\,mag brighter in $M_V$ and $M_K$ than their dM counterparts (Hawley et al.\ \cite{Hawley}). An estimate of the number of foreground dMe stars can be obtained by following the same approach as Dahm \& Simon (\cite{DS}), based on the stellar luminosity function of Jahrei\ss\ \& Wielen (\cite{JW}) and the dMe incidence of Hawley et al.\ (\cite{Hawley}). In this way, we arrive at an estimated number of 151 foreground dMe stars. We note that this number is an upper limit, since the reddening due to interstellar material along the line of sight towards NGC\,6383 was not taken into account in this evaluation. 

\begin{figure*}[htb]
\begin{minipage}{9cm}
\begin{center}
\resizebox{9cm}{!}{\includegraphics{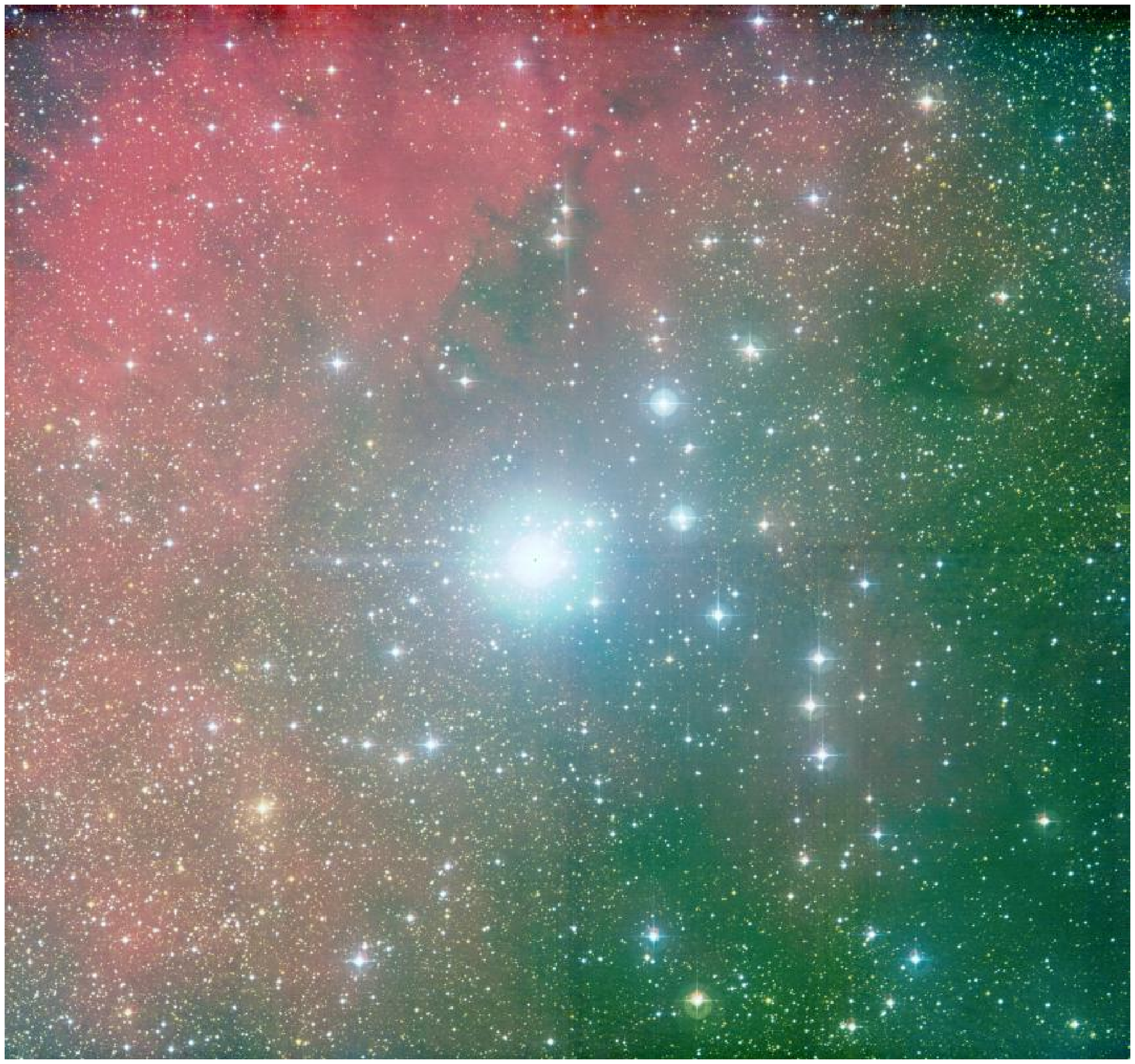}}
\end{center}
\end{minipage}
\hfill
\begin{minipage}{9cm}
\begin{center}
\resizebox{9cm}{!}{\includegraphics{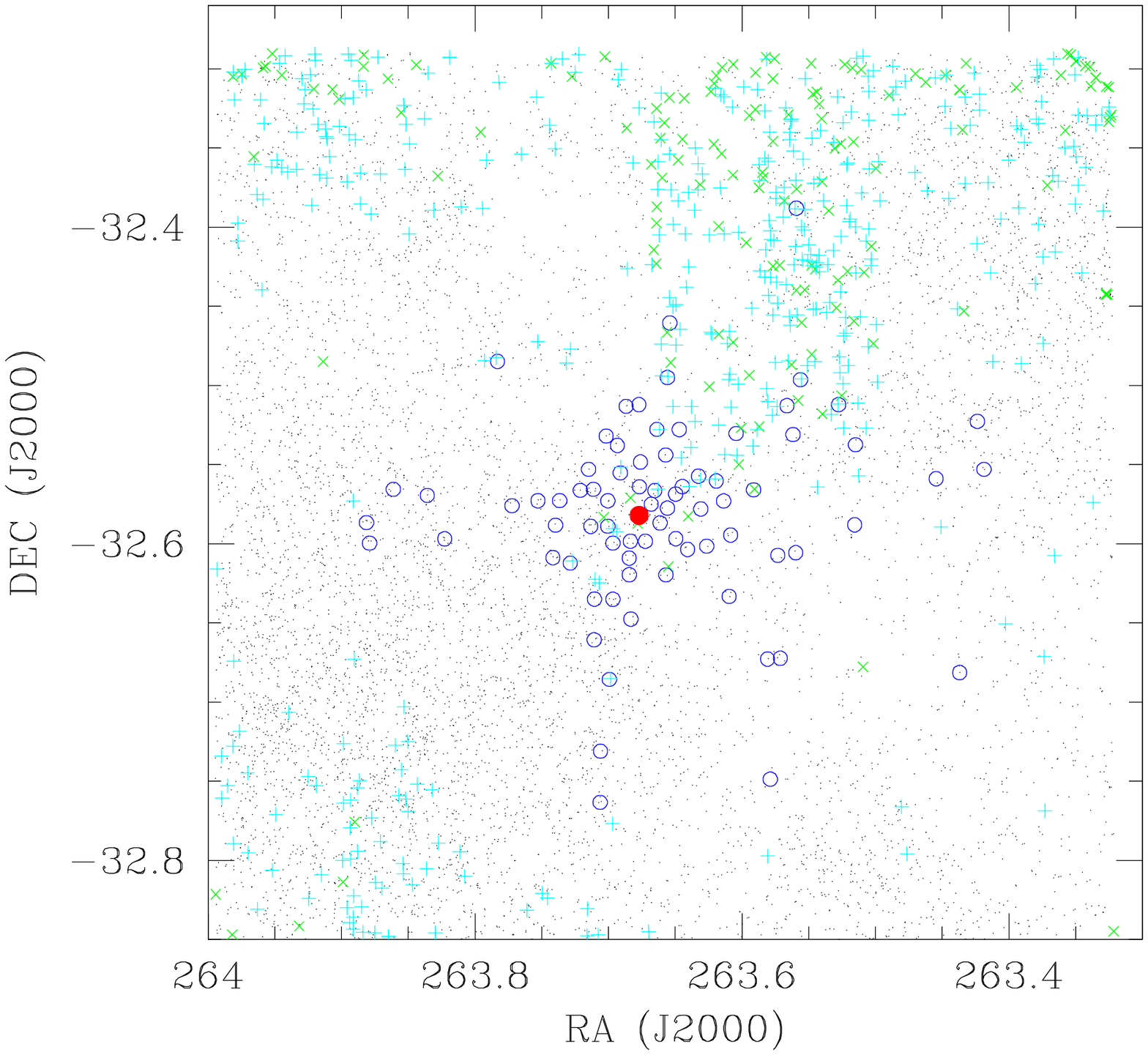}}
\end{center}
\end{minipage}
\caption{Left: three-colour image of the NGC\,6383 cluster compiled from our WFI photometry. Right: spatial distribution of the H$\alpha$ emitters ($\times$ signs) and candidates ($+$ signs). The large dot in the centre of the field indicates the position of HD\,159176, while the open circles represent the X-ray sources taken from Paper I. North is at the top and east is on the left.\label{spatial}}
\end{figure*}

A strong contamination by foreground dMe objects is {\it a priori} expected to produce a rather uniform spatial distribution of H$\alpha$ selected stars across the field of view. However, the spatial distribution of the H$\alpha$ emitters and candidates is far from uniform (see Fig.\,\ref{spatial}). Indeed, the density of emitters is higher towards the north of the field of view and especially in the north-western part, while there is a paucity of H$\alpha$ emitters in the south-west and to the east of HD\,159176. This can be quantified for instance by comparing the north-west and south-east quadrants: the former contains a total of 1951 objects with good photometric data, 18\% of which are either emitters or candidates, whereas the latter contains 3056 good quality objects only 3\% of which are selected as H$\alpha$ emitters or candidates. The lack of H$\alpha$ emitters and candidates in the south-east is particularly surprising because this region has the highest density of stars. Therefore, the paucity of emitters cannot be attributed to the effect of interstellar absorption. Although projection effects could play a role, we note that unlike our findings for X-ray emitters (see Paper I), there is no clear concentration of H$\alpha$ selected stars in the cluster core\footnote{Kharchenko et al.\ (\cite{Kharchenko}) estimate a core radius of 4.8\,arcmin and a corona radius of 15\,arcmin for NGC\,6383, while Piskunov et al.\ (\cite{Piskunov}) evaluate a tidal radius of at least 30\,arcmin.}. Outside the core, the mean spatial density of H$\alpha$ emitters and candidates are 0.08 and 0.26\,arcmin$^{-2}$, respectively. Taken together, these numbers are higher than expected from the predicted maximum number of foreground dMe stars (0.13\,arcmin$^{-2}$), indicating that the distribution of H$\alpha$ emitters and candidates must be related to the properties of the cluster or its background.

What is the reason for the lack of accreting PMS stars in the south-western and eastern parts of the halo of NGC\,6383, as inferred from the position of H$\alpha$ emitters and candidates? One possibility is photoevaporation of the protoplanetary disks by either the extreme-UV (EUV) and far-UV (FUV) irradiation from the O-type binary HD\,159176 or the EUV, FUV and X-rays of the central PMS star itself. The former effect was predicted theoretically, and evaporating disks with cometary tails were observed with {\it Spitzer} in three cases at distances between 0.1 and 0.35\,pc from an O-star (Balog et al.\ \cite{Balog}, \cite{Balog2} and references therein). However, in NGC\,6383, several H$\alpha$ emitters are relatively close to the O-star binary. We found 8 H$\alpha$ emitters and 19 candidates within 5\,armin of HD\,159176, which corresponds to a linear radius of 1.9\,pc at the distance of 1.3\,kpc. Therefore, we would have to assume that the photoevaporation occurs preferentially along certain directions and affects the cluster halo more efficiently than the core, which is rather unlikely. Concerning the second effect, Gorti \& Hollenbach (\cite{GH}) predict photoevaporation of the accretion disk by irradiation of the PMS star within a rather short timescale of about 1\,Myr. If the emission measure of the X-ray emitting plasma correlates with the EUV emission measure of the PMS star, this second scenario could, at least qualitatively, explain why the most prominent X-ray emission is observed from wTTs. This photoevaporation caused by irradiation by the central star itself should however affect PMS stars irrespective of their position in the cluster. This scenario alone is therefore insufficient to explain the complex distribution of H$\alpha$ emitters.

An alternative could be that our sample is heavily contaminated by background objects. For instance, we could be observing the edge of another star-forming region to the north of HD\,159176. We tested this possibility by considering the spatial distribution of objects with high $J - K$ indices ($J - K > 2.5$) that hence are potentially highly reddened background stars. This distribution reveals a lack of objects in the south-west, but is otherwise rather uniform. Therefore, while it is likely that our sample is highly contaminated by a population of background stars, we see no peculiar pattern in the spatial distribution of highly reddened stars.

\section{The Hertzsprung-Russell diagram}
From the various colour-magnitude diagrams presented in the previous section, we find that the X-ray sources and the Paunzen et al.\ (\cite{Paunzen}) PMS candidates (hereafter the PNZ objects) are the best candidate cluster members, whereas the H$\alpha$ emission criteria select a population that could contain a significant fraction of background objects as well as some field dMe stars.  

\begin{figure}[htb]
\begin{center}
\resizebox{9cm}{!}{\includegraphics{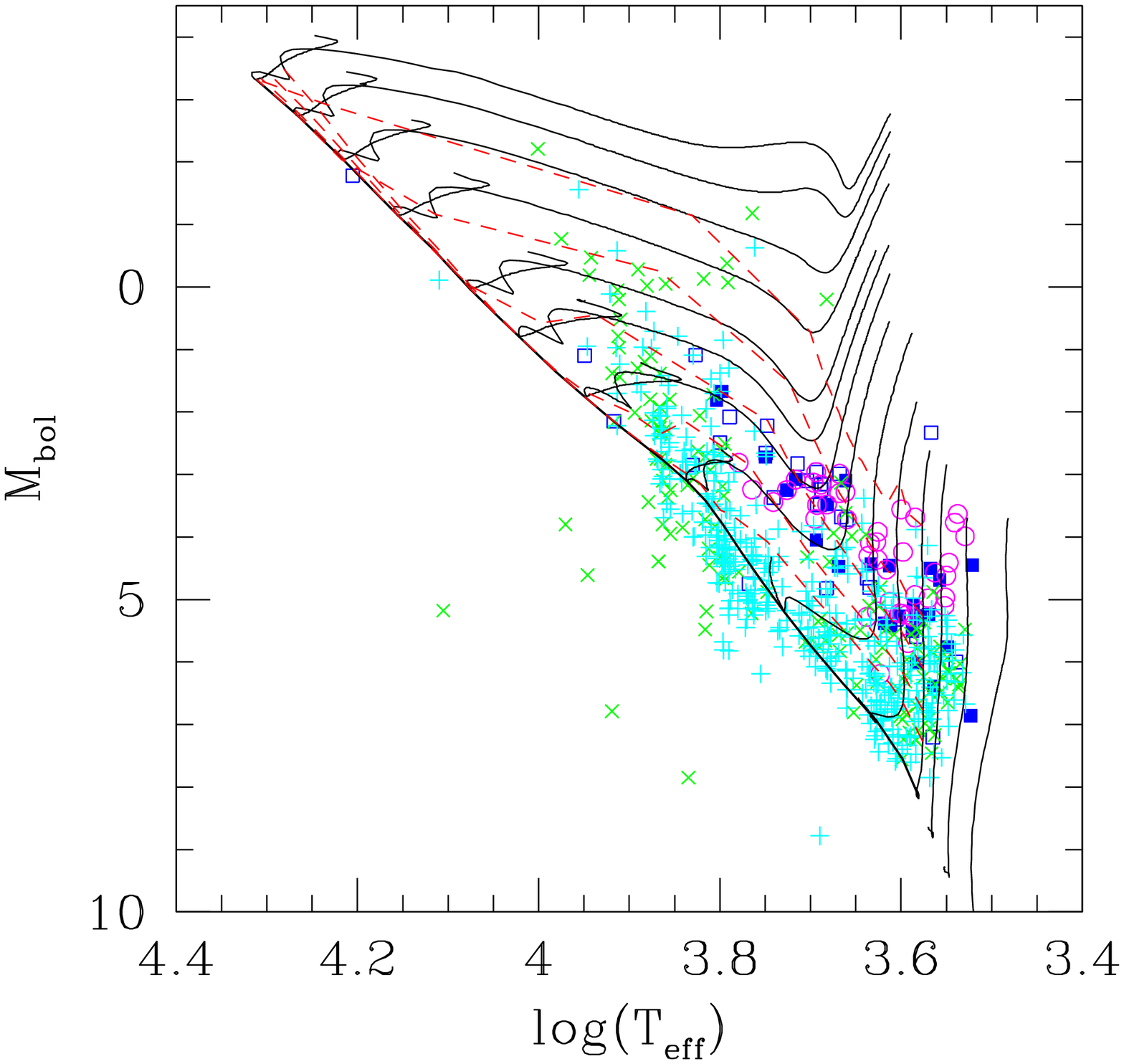}}
\end{center}
\begin{center}
\resizebox{9cm}{!}{\includegraphics{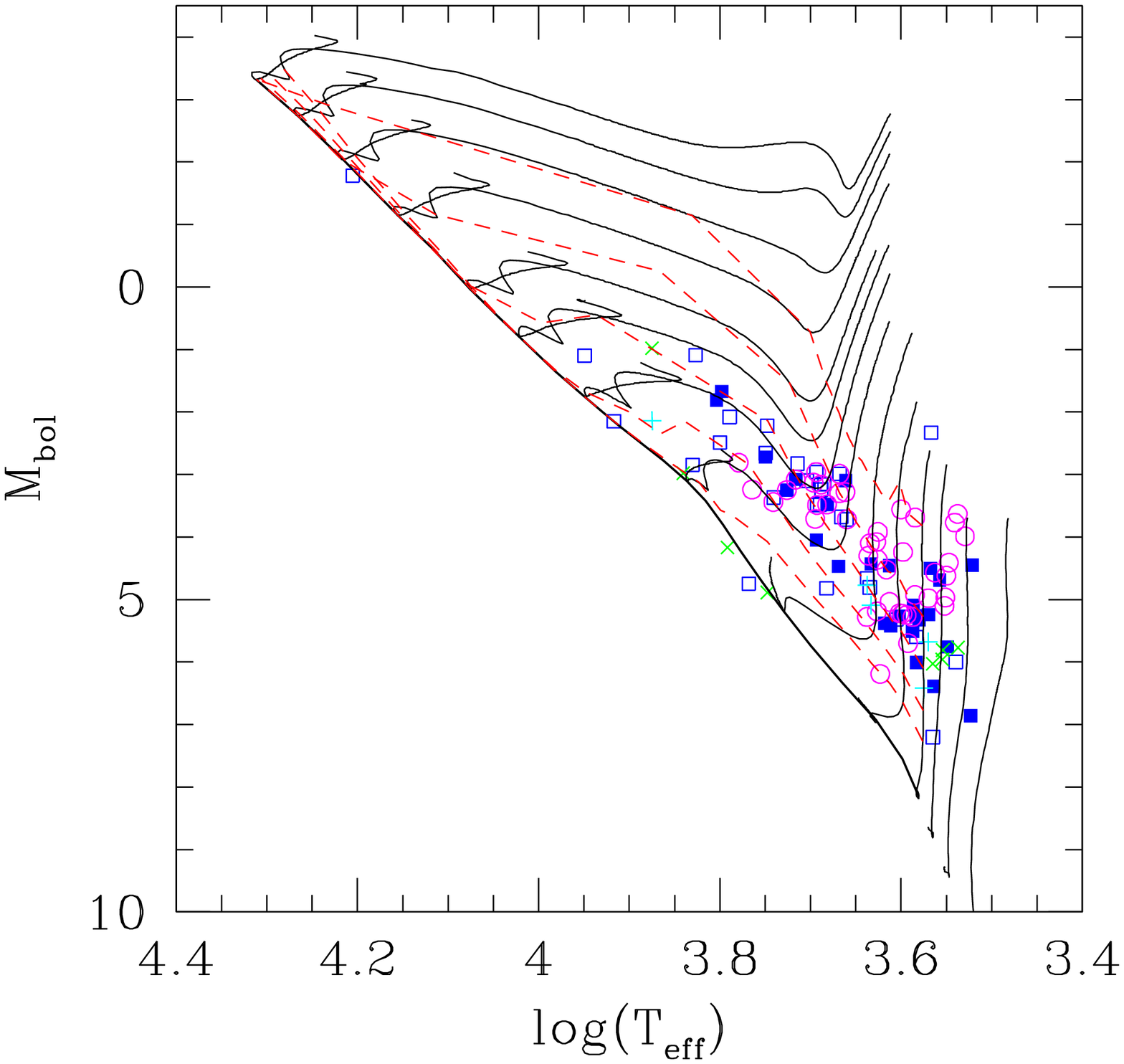}}
\end{center}
\caption{Top: Hertzsprung-Russell diagram of the X-ray sources, PNZ objects, H$\alpha$ emitters, and candidates. The PMS evolutionary tracks are taken from Siess et al.\ (\cite{Siess}) for masses of 0.2, 0.3, 0.4, 0.5, 0.7, 1.0, 1.5, 2.0, 2.5, 3.0, 4.0, 5.0, 6.0, and 7.0\,M$_{\odot}$. The thick solid line illustrates the zero-age main-sequence  and the dashed lines yield isochrones for ages of 0.5, 1.5, 4.0, 10.0, and 20.0\,Myr. The symbols have the same meaning as in Fig.\,\ref{figVIRHa}. Bottom: same, but with the selection criteria $J - K < 1.5$ and position within the 5\,arcmin radius of the cluster core applied to the H$\alpha$ emitters and candidates.\label{HRD}}
\end{figure}

Assuming a distance modulus of 10.57 and a reddening $E(B - V) = 0.32$ for the cluster members (see Rauw \& De Becker \cite{handbook} and Sect.\,\ref{intro}), we built the Hertzsprung-Russell diagram of the X-ray sources, the PNZ objects, and the H$\alpha$ selected stars. To do so, we evaluated the effective temperatures and bolometric corrections by means of a linear interpolation of the Kenyon \& Hartmann (\cite{KH}) tables as a function of $(V - I_c)_0$. The results are shown in Fig.\,\ref{HRD}. Whilst some of the H$\alpha$ selected objects form a continuation of the locus of the X-ray and PNZ objects, the majority of the H$\alpha$ emitters are located below this locus. 
This result must be caused by the contamination of our H$\alpha$ sample by 
foreground dMe stars or background objects located inside or behind the molecular cloud. Since our attempts to establish an efficient and self-consistent membership criterion failed, we instead applied a rough selection based upon the $J - K$ colour and the position with respect to the cluster core. By selecting only stars with $J - K < 1.5$, which is roughly equivalent to $A_K < 0.54$, and within the 5\,arcmin radius of the cluster core, we obtain a cleaner Hertzsprung-Russell diagram where the H$\alpha$ emitters are now mainly in a lower mass continuation of the locus of PNZ and X-ray selected PMS stars.\\

\begin{figure}[htb]
\begin{center}
\begin{minipage}{4.3cm}
\resizebox{4.4cm}{!}{\includegraphics{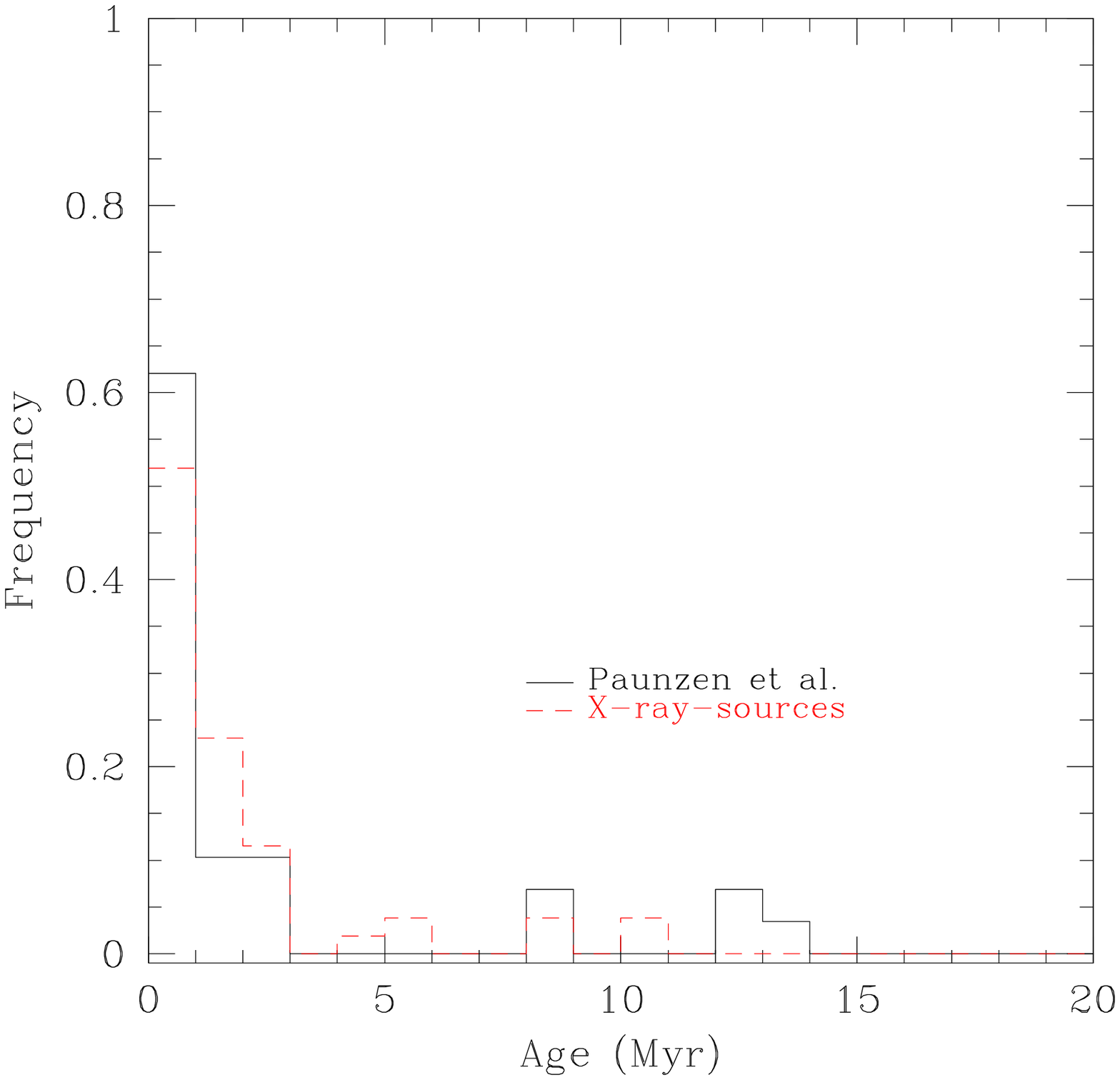}}
\end{minipage}
\begin{minipage}{4.3cm}
\resizebox{4.4cm}{!}{\includegraphics{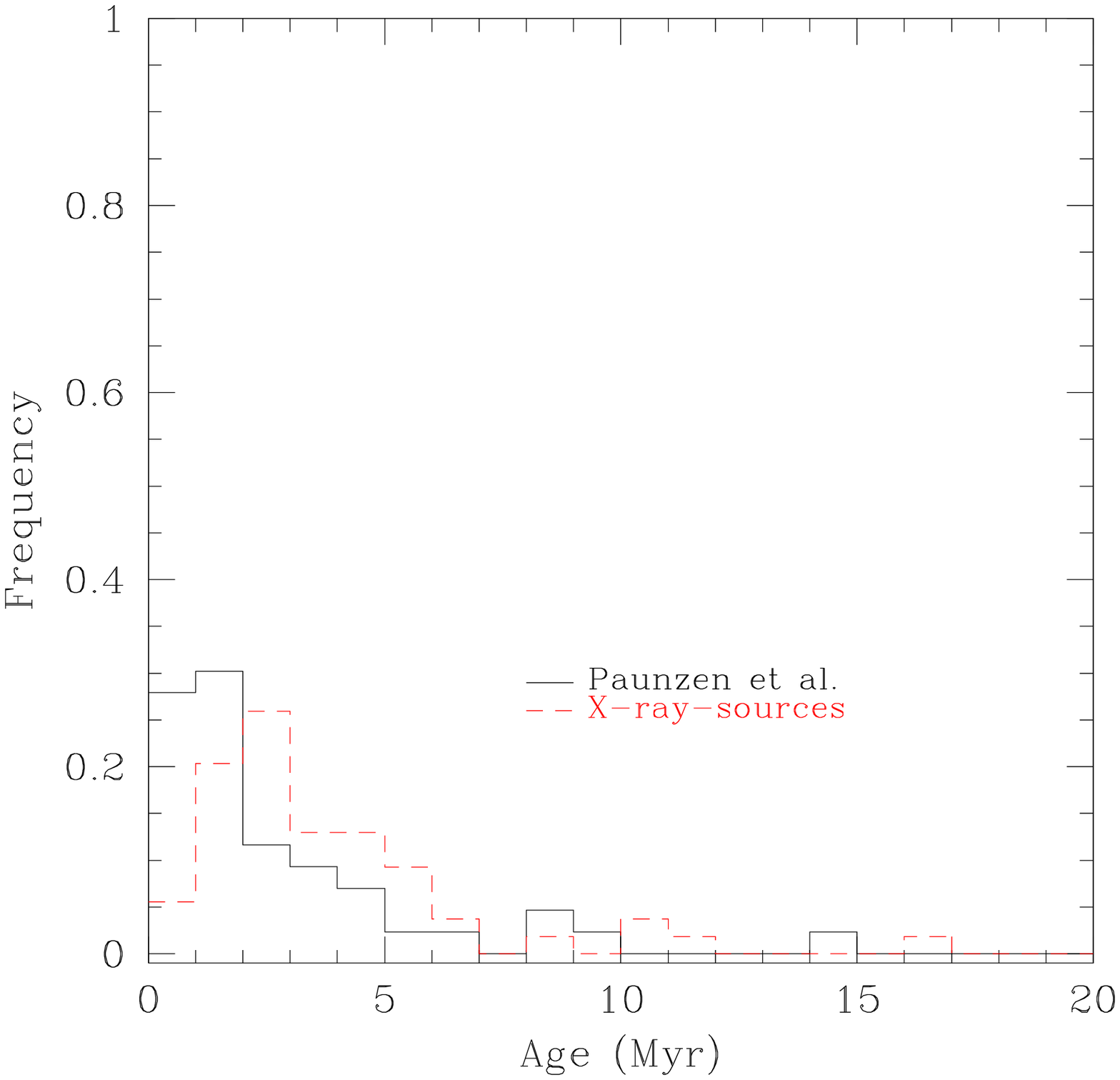}}
\end{minipage}
\end{center}
\begin{center}
\begin{minipage}{4.3cm}
\resizebox{4.4cm}{!}{\includegraphics{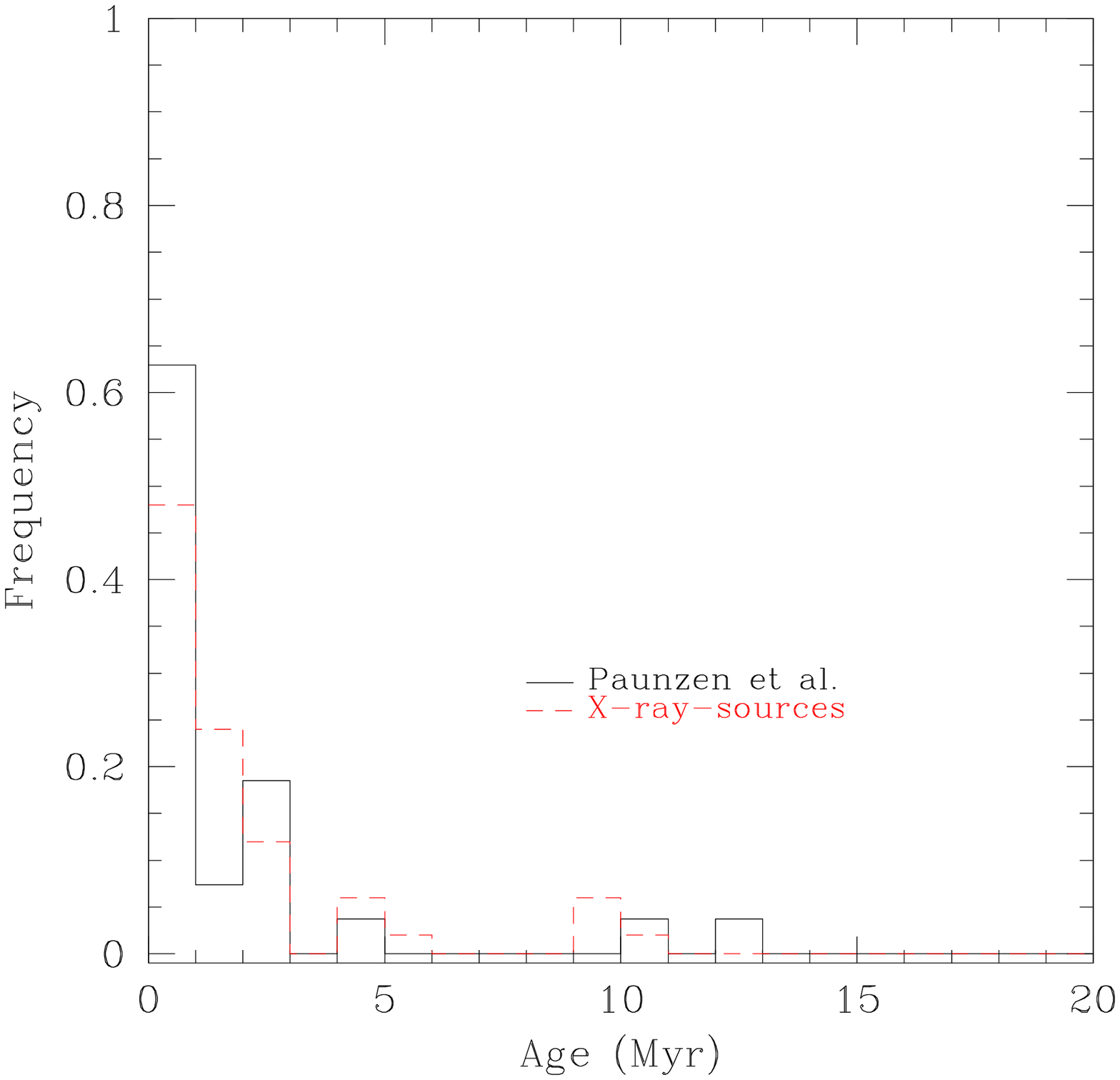}}
\end{minipage}
\begin{minipage}{4.3cm}
\resizebox{4.4cm}{!}{\includegraphics{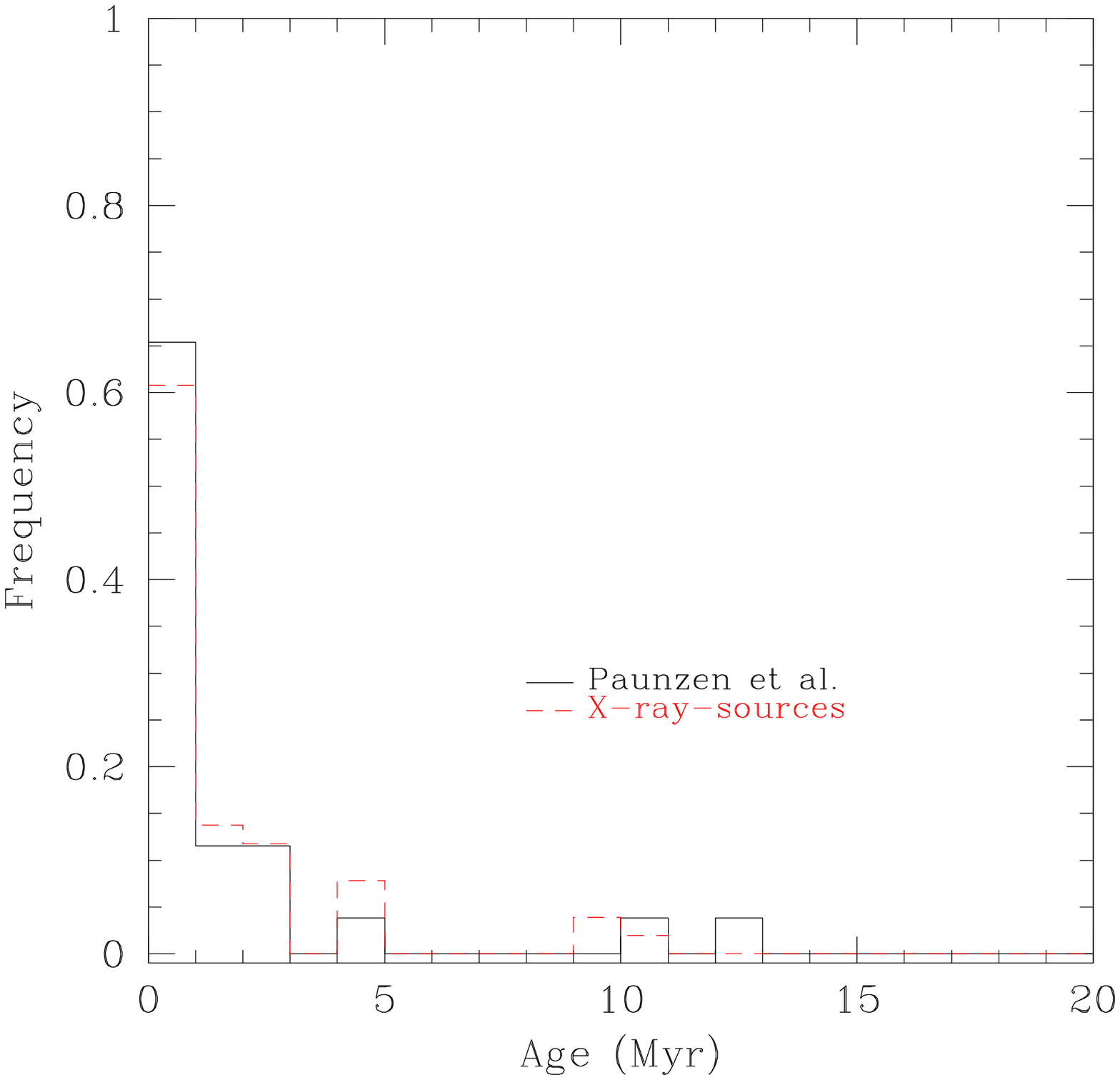}}
\end{minipage}
\end{center}
\caption{Distribution of ages as inferred from the comparison with PMS isochrones. The evolutionary models are from D'Antona \& Mazzitelli (\cite{DM97}, top left), Siess et al.\ (\cite{Siess}, top right), and Landin et al.\ (\cite{Landin}) without rotation (lower left) and with rotation (lower right). \label{ages}}
\end{figure}

There are a number of available theoretical PMS tracks that we can compare with our data. The various evolutionary codes differ in many respects and the most important differences concern the opacities, the equation of state, the treatment of the stellar atmosphere, and most importantly the treatment of convection (see the discussions in Siess et al.\ \cite{Siess}). Here, we chose to compare our data with the results of four different models: the 1997 version of the tracks of D'Antona \& Mazzitelli (\cite{DM97}, hereafter DM97), the Siess et al.\ (\cite{Siess}, hereafter SDF00) calculations, and both the standard and rotating, models of Landin et al.\ (\cite{Landin}, hereafter LMV09). Whilst the DM97 computations use the full spectrum of turbulence (FTS) approach to treat convection, the other calculations considered here rely instead on the mixing length theory (MLT). These different treatments obviously lead to quantitative differences in the PMS tracks, and one expects conflicting results when empirically determining the ages of observed PMS stars. For instance, D'Antona \& Mazitelli (\cite{DM97}) argue that the FTS approach generally yields a lower age spread than the MLT. With these limitations in mind, we compiled the histograms of the ages of the PMS candidates by comparing with the isochrones of the various models (see the results in Fig.\,\ref{ages} and Table\,\ref{tabages}). We emphasize that the age distribution of the H$\alpha$ selected stars would actually be meaningless because, as stated above, this selection criterion probably fails to properly identify cluster members.
\begin{table}[h]
\caption{Mean ages and standard age deviations (in Myr) for the various samples of PMS candidates in NGC\,6383.\label{tabages}}
\begin{tabular}{l c c c c}
\hline
& \multicolumn{2}{c}{X-ray selected} & \multicolumn{2}{c}{PNZ objects} \\
& mean & stand.\ dev. & mean & stand.\ dev. \\
\hline
DM97       & 1.9 & 2.5 & 2.6 & 4.0 \\
SDF00      & 3.8 & 3.0 & 2.8 & 2.9 \\
LMV09 std. & 2.1 & 2.6 & 1.9 & 2.9 \\
LMV09 rot. & 1.7 & 2.3 & 1.8 & 3.0 \\
\hline
\end{tabular}
\end{table}

Figure\,\ref{ages} shows that the DM97 isochrones indeed yield a strongly peaked age distribution (especially for the PNZ objects). We note that this information is not obvious from the standard deviations listed in Table\,\ref{tabages}, since the few outliers at ages of 10\,Myr and older have a strong impact on these parameters. Although the quantitative picture depends upon the evolutionary models in use, we note that the X-ray sources and the PNZ objects have essentially identical age distributions which are strongly peaked below 5\,Myr and the likely PMS stars have a mean age between 2 and 3\,Myr. From the observational point of view, the combined uncertainty in cluster distance and reddening (as inferred for the brighter cluster members) amounts to 0.18\,mag, which introduces an uncertainty of about 0.4\,Myr in the cluster age. The latter number is well within the dispersion of the ages implied from the comparison with the different evolutionary models.

Our results are in reasonable agreement with the upper age limit of 4\,Myr inferred by Paunzen et al.\ (\cite{Paunzen}). We note that FitzGerald et al.\,(\cite{Fitz}) derived a younger cluster age of $1.7 \pm 0.4$\,Myr. The difference may originate from the use of different PMS isochrones or the work of FitzGerald et al.\,(\cite{Fitz}) being restricted to the brighter cluster members only.

Finally, we estimated the age of the central early-type binary HD\,159176 from the effective temperature and bolometric luminosity determined by the spectral analysis of Linder et al.\ (\cite{Linder}). The bolometric luminosities were computed by assuming that $A_V = 0.96$ and $DM = 10.57$, while the effective temperatures were adapted from Martins et al.\ (\cite{Martins}) following the spectral types derived by Linder et al.\ (\cite{Linder}). Since HD\,159176 is probably a detached binary system, binary evolution effects are unlikely to have played a major role in the evolution of the system to date. Therefore, we compared the position of the stars in the Hertzsprung-Russell diagram with the single-star evolutionary tracks of Meynet \& Maeder (\cite{MM}). Depending on whether we use the evolutionary tracks with an initial equatorial rotational velocity of 0 or 300\,km\,s$^{-1}$, we estimate the age of the system to be in the range 2.3 -- 2.7\,Myr, in good agreement with the $2.8 \pm 0.5$\,Myr estimated by FitzGerald et al.\ (\cite{Fitz}). As a result, we find that the age of the massive binary and the mean age of the PMS stars are reasonably consistent with a single star-formation event about 2 -- 3\,Myr ago.\\ 

We emphasize that many objects meet more than one PMS criterion simultaneously. This is the case for the X-ray sources \#16, 20, 25, 39, 42, 47, 51, 57, and 63 that appear in the list of PMS candidates of Paunzen et al.\ (\cite{Paunzen}). Five of these objects were observed spectroscopically and were found to exhibit a significant absorption probably caused by Li\,{\sc i} $\lambda$\,6708. For three of them, we also found weak H$\alpha$ emission (\#20, 25 and 57) in the spectra. Finally, we note that our X-ray source \#51 is identical with WEBDA object 382, which is understood to be a rapidly rotating PMS star (Paunzen et al.\ \cite{Paunzen}, Zwintz et al.\ \cite{Zwintz}, their object 71). 

\section{Conclusions}
Our photometric and spectroscopic investigation of NGC\,6383 has detected a significant population of reddened objects with extinctions spanning a continuous range up to $A_V \simeq 20$. We suggest that these objects are located at different depths within the natal molecular cloud of the cluster. NGC\,6383 could therefore be the visible part of a significantly larger star formation event, probably affecting large parts of this molecular cloud. 

We have found that the best PMS candidates display little or no H$\alpha$ emission in their spectrum. The mean age of the most likely low-mass PMS candidates in NGC\,6383 is found by comparing with various PMS evolutionary tracks, to be 2 -- 3\,Myr, in reasonable agreement with the estimated age of HD\,159176 at the centre of the cluster. It therefore appears that the low-mass stars and the O-type binary were born in the same star-formation event. 

\acknowledgement{We acknowledge support from the Fonds de Recherche Scientifique (FRS/FNRS), through the XMM/INTEGRAL PRODEX contract as well as by the Communaut\'e Fran\c caise de Belgique - Action de recherche concert\'ee - Acad\'emie Wallonie - Europe. This research made use of data products from the Two Micron All Sky Survey, which is a joint project of the University of Massachusetts and the Infrared Processing and Analysis Center/California Institute of Technology, funded by NASA and NSF.}

\appendix
\section{Notes on individual stars\label{app}}
Star \#1: the strength of the G-band and the ratio Fe\,{\sc i} $\lambda$\,4325/H$\gamma$ suggest a K0 spectral type. The strength of Y\,{\sc ii} $\lambda$\,4376 relative to Fe\,{\sc i} $\lambda$\,4383 and the general appearance of the violet-system CN bands and Ca\,{\sc ii} H and K lines indicate a main-sequence or slightly brighter luminosity class. We thus adopt a K0\,V spectral type with an uncertainty of about two spectral subtypes. 

Star \#3: the strength of the H8 line compared to the Ca\,{\sc ii} H and K lines, the strength of the (weak) G-band, and the relative importance of metallic line spectrum indicate an F5 spectral-type. The weakness of the Sr\,{\sc ii} $\lambda$\,4077 line indicates a main-sequence or subgiant luminosity class. We thus classify this star as F5\,V with an uncertainty of two spectral subtypes.

Star \#16: the strength of the G-band and the Ca\,{\sc i} $\lambda$\,4227 line as well as the ratio Fe\,{\sc i} $\lambda$\,4325/H$\gamma$ suggest an early G (G0-5) spectral type. The strength of Sr\,{\sc ii} $\lambda$\,4077 suggests either a subgiant or giant luminosity class. The strength of Y\,{\sc ii} $\lambda$\,4376 relative to Fe\,{\sc i} $\lambda$\,4383 is instead indicative of a main-sequence luminosity class. We thus adopt a G0-5\,V-IV spectral type.

Star \#20: the strength of the G-band and the ratio Fe\,{\sc i} $\lambda$\,4325/H$\gamma$ yield a spectral type of about K0. The strength of Sr\,{\sc ii} $\lambda$\,4077 compared to the nearby Fe\,{\sc i} lines suggests a main-sequence or slightly brighter luminosity class. We thus classify this star as K0\,V with two subclasses of uncertainty. The star exhibits a rather strong H$\alpha$ emission above the continuum with an equivalent width of about $-1.0$\,\AA.

Star \#25: the same criteria as for star \#20 yield an early to mid K spectral type. The strength of Sr\,{\sc ii} $\lambda$\,4077 compared to the nearby Fe\,{\sc i} lines suggests a main-sequence luminosity class. This is also in agreement with the strength of Y\,{\sc ii} $\lambda$\,4376 relative to Fe\,{\sc i} $\lambda$\,4383. We thus adopt a K2\,V spectral type with an uncertainty of about two spectral subtypes. The H$\alpha$ line is seen in emission above the continuum with an EW of $-0.5$\,\AA.

Star \#39: the strength of the G-band indicates that this star must be earlier than K5, whilst the weakness of the Balmer lines and the ratio of Fe\,{\sc i} $\lambda$\,4325/H$\gamma$ suggest a spectral type later than K0. Considering also the general appearance of the spectrum, we favour a spectral type K2-4. The same criteria as for star \#25 indicate a main-sequence luminosity class. We thus adopt a K2-4\,V spectral type. While no H$\alpha$ emission is seen above the continuum, there is also no absorption at the wavelength of H$\alpha$. There is an absorption towards the blue, suggesting a kind of P-Cygni profile.

Star \#50: the strength of the Ca\,{\sc i} $\lambda$\,4227 line and the MgH feature at 4780\,\AA\ suggest a late-K main-sequence type, whilst the lack of visible TiO bands indicates that the star must be earlier than M0. We adopt a K7\,V spectral type with an uncertainty of at least two subtypes. A strong (EW = $-3.0$\,\AA) H$\alpha$ emission is detected.

Star \#54: the strength of the G-band and the Ca\,{\sc i} $\lambda$\,4277 line as well as the ratio Fe\,{\sc i} $\lambda$\,4325/H$\gamma$ suggest a late G spectral type. The strength of Sr\,{\sc ii} $\lambda$\,4077 as well as the strength of Y\,{\sc ii} $\lambda$\,4376 compared to Fe\,{\sc i} $\lambda$\,4383 yield a main-sequence or giant luminosity class. We thus adopt a G7\,V spectral type with an uncertainty of two subtypes in spectral type. 

Star \#55: the same criteria as for star \#54 yield an early-G type, G0-5, but closer to G0. The strength of the Ti\,{\sc ii}, Fe\,{\sc ii} $\lambda\lambda$ 4172-78 blend, as well as of the Sr\,{\sc ii} $\lambda$\,4077 line suggest a subgiant luminosity class. We classify this star accordingly as G0-1\,IV. 

Star \#56: as inferred by the presence of He\,{\sc i} lines, this object is a B-star. The intensity ratio He\,{\sc i} $\lambda$\,4471 versus Mg\,{\sc ii} $\lambda$\,4481 as well as the strength of the Ca\,{\sc ii} K line indicate a spectral type B5-7. The rather broad Balmer lines suggest a main-sequence luminosity class. We thus classify this star as B5-7\,V. The spectrum also exhibits the diffuse interstellar band (DIB) at 4430\,\AA. 

Star \#57: the same criteria as for star \#39 suggest a spectral type K0-5. The strength of Sr\,{\sc ii} $\lambda$\,4077 compared to the nearby Fe\,{\sc i} lines suggests a giant luminosity class, whilst the strength of Y\,{\sc ii} $\lambda$\,4376 compared to Fe\,{\sc i} $\lambda$\,4383 could be consistent with either a main-sequence or giant classification. Considering also the general appearance of the spectrum, we adopt a K0-5\,III classification. Our spectrum displays a blue-shifted H$\alpha$ emission above the continuum with an equivalent width of about $-0.3$\,\AA. 

Star \#65: the same criteria as for star \#16 indicate a G5 spectral type. The strength of the Sr\,{\sc ii} $\lambda$\,4077 line suggests that the star is a subgiant or giant. We thus classify this star as G5\,IV-III.

Star \#67: the same criteria as for stars \#16 and \#25 yield a G5-K0\,V spectral type.

Star \#70: the same criteria as for star \#16 indicate a late G spectral type. The strength of Sr\,{\sc ii} $\lambda$\,4077 compared to the nearby Fe\,{\sc i} lines and the strength of Y\,{\sc ii} $\lambda$\,4376 relative to Fe\,{\sc i} $\lambda$\,4383 suggest a subgiant or giant luminosity class. We thus classify the star as G5-K0\,IV-III.

Star \#76: the spectrum of this source is very red with almost no flux below $\sim 4300$\,\AA. However, this must be a rather hot object, since in addition to some Balmer absorption lines, we also see several He\,{\sc i} absorption lines ($\lambda\lambda$ 4471, 4922, 5876, 6678). The Mg\,{\sc ii} $\lambda$\,4481 line is weaker than He\,{\sc i} $\lambda$\,4471. The spectrum also exhibits a number of strong diffuse interstellar bands. This star must therefore be a heavily extinguished object of spectral type O9-B5. 

\end{document}